# Interaktion mit 3D-Objekten in Augmented Reality Anwendungen auf mobilen Android Geräten

Abschlussarbeit

zur Erlangung des akademischen Grades

**Bachelor of Science (B.Sc.)**

an der

Hochschule für Technik und Wirtschaft Berlin

Fachbereich Wirtschaftswissenschaften II

Studiengang Angewandte Informatik

eingereicht von:     Lennart Brüggemann

Matrikelnummer:     525078

1. Prüfer:     Prof. Dr.-Ing. Thomas Jung

2. Prüfer:     M. Eng. Thomas Damerau

Datum:     14. Februar 2012

In Zusammenarbeit mit dem Fraunhofer Institut für Produktionsanlangen und Konstruktionstechnik (IPK)

# Zusammenfassung

Diese Bachelorarbeit beschreibt den Entwurf und die Implementierung einer Augmented Reality Anwendung für die Android Plattform. Ziel der Anwendung ist es, die Interaktionsmöglichkeiten in einer Augmented Reality Umgebung auf mobilen Geräten zu demonstrieren.

Dazu wird ein 3D-Modell mittels markerbasiertem Tracking auf einem Touchscreen dargestellt, das der Benutzer nach Belieben verschieben, rotieren und skalieren kann. Ferner kann er einzelne, vorher festgelegte Einzelteile des Modells anwählen und hervorheben, und sich kontextsensitiv Informationen zu diesen Teilen ansehen.

Um Entwicklern die einfache Modifizierbarkeit der Anwendung zu ermöglichen, werden die Informationen in einen eigenen Datentyp ausgelagert, um sie ohne größere Änderungen des Quellcodes an sich ändernde Anforderungen anpassen zu können.

Nach einer Einführung in Augmented Reality, der ihr zugrundeliegenden Technologie und die Android-Plattform, werden Einsatzmöglichkeiten für die Applikation vorgestellt und deren benötigte Funktionalität beschrieben. Anschließend wird ein Entwurf der Anwendung erstellt und die letztendliche Implementierung erklärt.

# Abstract

This bachelor's thesis describes the conception and implementation of an augmented reality application for the Android platform. The intention is to demonstrate some possibilities of interaction within an augmented reality environment on mobile devices.

For that purpose, a 3D-model is displayed on the devices' touchscreen using marker-based tracking. This enables the user to translate, rotate or scale the model as he wishes. He can additionally select and highlight preassigned parts of the model to display specific information for that element.

To assist developers in modifying the application for changing requirements without re-writing large portions of the source code, the information for each part have been encapsulated into its own data type.

After an introduction to augmented reality, its underlying technology and the Android platform, some possible usage scenarios and the resulting functionalities are outlined. Finally, the design as well as the developed implementation are described.

# Inhaltsverzeichnis

# 1. Einleitung

## 1.1. Motivation

Durch die immer größere Verbreitung von Smartphones und Tablet-PCs sind tragbare Computer im Kleinformat mittlerweile nicht mehr nur zum Telefonieren, Spielen und Arbeiten interessant. Neben den Anwendungsgebieten im Alltag, können kompakte Computer mit Touchscreens auch in der Industrie eingesetzt werden.

Bei der Planung und Erstellung von Produkten kann ein digitales Mock-Up (DMU), also ein computergeneriertes Modell des sich in Entwicklung befindlichen Produkts, durch virtuelle Bauraum- oder Kollisionsanalysen Zeit und Kosten sparen, und leicht an sich ändernde Anforderungen angepasst werden.

DMUs können bereits mit CAD-Programmen an Computern betrachtet und leicht manipuliert werden, und damit die Anzahl von aufwendig herzustellenden physischen Mock-Ups (PMU) reduzieren [KFG07_a]. Können bei realen Modellen meist noch relativ einfach kleinere Modifikationen des Modells durchgeführt werden, ist der Austausch größerer Teile (wie etwa der Frontpartie eines Autos) schon wesentlich aufwendiger – sofern nicht sogar ein komplett neues Modell erstellt werden muss. Allerdings kann ein DMU ein physisches Modell in Hinblick auf realistische Maßstäbe und visuelle Wahrnehmung nicht ersetzen.

Daher wäre es wünschenswert, wenn sich ein Entwickler ein DMU einerseits in realen Proportionen ansehen, und gleichzeitig Änderungen daran vornehmen oder alternative Versionen (z. B. Farben) betrachten könnte.

Laut einer Studie des Fraunhofer IPT [KSP03] findet Augmented Reality (siehe 2.1) im Gegensatz zu Virtueller Realität (siehe 2.1.1) in der Industrie nur wenig Anwendung, wobei AR als noch zu wenig effizient angesehen wird (ebd. S. 7) und VR größtenteils für Entwurfsprüfungen und DMUs genutzt wird (ebd. S. 12). Mehr als die Hälfte der Befragten erwarten zudem eine vermehrte Nutzung von mobilen VR-Systemen in der Zukunft (ebd. S. 33). Ein mobiles System, das sich für den Einsatz in Entwurfsbesprechungen und die Darstellung von DMUs eignet, und die effektive Einsetzbarkeit von AR demonstriert, ist daher die Motivation für diese Arbeit.

## 1.2. Zielsetzung

Das Ziel dieser Arbeit ist die Erstellung eines Konzeptes und die Umsetzung einer mobilen Anwendung für die Android Plattform, mit der die Vorteile eines virtuellen und physischen Produktprototypen kombiniert werden sollen. Ein einfach zu veränderndes und schon vorhandenes CAD-Modell soll so dargestellt werden, dass ein realistischer Größeneindruck gewonnen werden kann und ohne aufwendige Nachmodellierung reversibel in begrenztem Maße verändert werden kann.

Der Benutzer soll mittels eines Smartphones oder Tablet-PCs das CAD-Modell eines Produktes in einem möglichst realistischen Maßstab betrachten können, und über eine geeignete Benutzerschnittstelle Transformationen (Rotation, Translation, Skalierung) darauf anwenden können.

Um die Anforderungen an die Implementierung zu ermitteln, werden außerdem einige mögliche Anwendungsfälle anhand des unter Problemstellung erwähnten Beispiels entwickelt.

Die Darstellung erfolgt über Augmented Reality. Das heißt, dass der Benutzer die Kamera seines mobilen Gerätes auf einen Marker (eine Art Bild, das die Software erkennen kann) richten kann, und das Modell dann im Kamerabild auf dem Marker erscheint.

## 1.3. Aufbau der Arbeit

Nach der Einleitung in diesem Kapitel, werden in Kapitel 2 die Grundlagen beschrieben, die für das Verständnis der Arbeit notwendig sind. Danach wird in Kapitel 3 ein kurzer Überblick über den Stand der Technik im Bereich der Augmented Reality Technologien gegeben.

In Kapitel 4 werden die Anforderungen an die Implementierung der Android-Anwendung erarbeitet und hinsichtlich der Qualitätsmerkmale analysiert. Außerdem werden einige mögliche Einsatzzwecke der Anwendung beschrieben und die Anwendungsfälle für die spätere Implementierung herausgestellt.

Die Anforderungen werden in Kapitel 5 in einen konkreten Entwurf überführt. Dazu gehören die Auswahl der Zielplattform und benötigter Drittkomponenten. Hier werden außerdem die Bestandteile der zu implementierenden Anwendung konzipiert.

Basierend auf den Anforderungen und Entwürfen aus den beiden vorangegangen Kapiteln, wird in Kapitel 6 die Umsetzung der vorliegenden Implementierung erklärt. Dieses beinhaltet auch einige Quellcode-Ausschnitte aus den Funktionen, die die realisierten Anwendungsfälle enthalten.

In Kapitel 7 wird ein Fazit gezogen und die Arbeit zusammengefasst. Außerdem werden Limitationen der Anwendung erläutert und einige denkbare Erweiterungsmöglichkeiten beschrieben.

# 2. Grundlagen

Dieses Kapitel soll einen Überblick über die in dieser Arbeit verwendete Terminologie geben und in die Technologien einführen, die die Grundlage der Arbeit bilden.

Begonnen wird mit einer Einführung in Augmented Reality. Da die Objektverfolgung ein integraler Bestandteil von Augmented Reality ist, werden außerdem verschiedene Ansätze aus dem Bereich der Computer Vision (CV, deutsch: „*maschinelles Sehen*") vorgestellt, die dies ermöglichen.

Danach wird die Architektur der Android-Plattform erläutert, und welche Möglichkeiten der Interaktion im Zusammenhang mit mobilen Geräten unterstützt werden.

## 2.1. Augmented Reality

Augmented Reality (AR, deutsch: „*angereicherte Realität*") bezeichnet die Darstellung virtueller Objekte über der realen Welt. Dabei werden die in der Regel computergenerierte Objekte (Informationen, 3D-Modelle, Grafiken) entweder über dem Kamerabild eines Gerätes (Mobiltelefon, Webcam) auf einem Bildschirm angezeigt, oder auf halbtransparente Oberflächen projiziert, durch die der Benutzer teilweise hindurchsehen kann (siehe auch 2.1.2). Azuma präzisiert in [Azuma97] zudem, dass AR die Realität ergänze, und nicht vollständig ersetze[1].

---

[1]Original: „*Therefore, AR supplements reality, rather than completely replacing it.*"

Da in den letzten Jahren, durch die breite Verfügbarkeit und vergleichsweise günstigen Preise, leistungsstarke mobile Geräte mit integrierten Kameras weit verbreitet sind, findet AR aktuell verstärkt im mobilen Sektor Anwendung. Touristen können ihr Smartphone auf das Brandenburger Tor richten und sich mit entsprechenden Anwendungen historische Daten, Videos oder auch Restaurants in der Nähe per AR über dem Kamerabild anzeigen lassenö. Die passenden Informationen werden über ortsabhängige Dienste (engl. *location-based services*, siehe auch 2.5.5) gefunden und zugeordnet. Dies ist z. B. mit dem Wikitude[2] AR Browser möglich.

Allerdings lässt sich AR auch für wissenschaftliche und industrielle Zwecke nutzen. In der Medizin können z. B. digitale Bilddaten aus der Radiologie über das Videobild von Endoskopiekameras gelegt werden[3], und so den Chirurgen bei Operationen in Echtzeit mit Informationen versorgen („Wie weit liegt die Arterie vom Schnitt entfernt?"). Am Heinz Nixdorf Institut der Universität Paderborn wird in Zusammenarbeit mit der Volkswagen AG eine AR-Versuchsplattform entwickelt [HNI], auf der die Benutzer über eine 3D-Datenbrille das Innere eines Fahrzeugs betrachten kann, um so Aussagen über die Ergonomie des Fahrzeuginnenraumen machen zu können.

Mit Spatial AR (SAR, deutsch: *„räumliche angereicherte Realität"*) können zudem mit einem Projektor Bilddaten auf ein physisches Objekt projiziert werden. So ist es möglich, die optischen Eigenschaften eines Objektes zu variieren, und z. B. eine Metall- oder Marmortextur auf eine Vase mit einer neutralen, weißen Oberfläche zu projizieren [BR05].

## 2.1.1. Virtuelle Realität

Zu AR abzugrenzen ist die *Virtuelle Realität* (VR). Im Gegensatz zu AR ist sie immersiv, d. h. sie umgibt den Anwender und versetzt ihn in eine Welt, die er als real wahrnimmt. Eine CAVE (*Cave Automatic Virtual Environment*), in der mehrere Projektoren in einem Raum eine dreidimensional wirkende Umgebung auf die Wände projizieren, ist ein Beispiel für virtuelle Realität.

---

[2]`http://www.wikitude.com/en/`
[3]Beispielsweise am Deutschen Krebsforschungszentrum, siehe Dkfz.

## 2.1.2. Head Mounted Displays

Ein weiteres Beispiel für die Anwendung von Augmented und Virtual Reality sind *Head Mounted Displays* (HMD). Dabei sind zwei kleine Bildschirme und optional eine Kamera auf einem Gestell fixiert, das sich der Benutzer auf den Kopf schnallen kann. Nach [RHF94] gibt es zwei mögliche Bauweisen für HMDs. Bei der Ersten wird das Blickfeld des Anwenders durch das HMD vollständig abgedeckt, und er blickt auf zwei kleine Bildschirme, die stereoskopisch den Blick in eine virtuelle Welt simulieren. Dieses Konzept entspricht der virtuellen Realität, da der Benutzer vollständig von der Außenwelt abgeschirmt ist. Bei der anderen Möglichkeit blickt der Benutzer durch eine Art Brille, auf die mittels eines integrierten Projektors Informationen projiziert werden. Da der Anwender dadurch sowohl die reale Welt, als auch virtuelle Objekte darüber sieht, entspricht dies AR.

Beide Möglichkeiten können durch eine Kamera ergänzt werden, die den ungefähren Blickwinkel des Benutzers aufnimmt. Mittels Objekt-Erkennung und -Verfolgung können, entsprechend der gefundenen Objekte, dann Daten in das Blickfeld des Benutzers gerendert werden, dies entspricht ebenfalls AR.

In der Industrie kommen HMDs beispielsweise zur Reparaturhilfe bei Kraftfahrzeugen zum Einsatz [BGW11]. Daneben finden HMDs etwa in der Militärtechnik Anwendung. Als *Helmet* Mounted Displays sind sie in die Helmen von Piloten integriert, um sie während des Fluges mit Informationen zu versorgen und bei der Zielfindung zu assistieren [EF].

## 2.2. Objekterkennung und -Verfolgung

Unabhängig von der Anwendung ist die Erkennung und Verfolgung von Objekten (engl. *Tracking*) ein wichtiger Teil der AR. Die benötigte Präzision und Stabilität schwankt je nach Anwendungsfall: Beim Brandenburger Tor kommt es nicht auf millimetergenaue Erkennung an, in der Chirurgie kann ein Millimeter jedoch lebensentscheidend sein.

### 2.2.1. Markerbasiertes Tracking

Eine Möglichkeit der Objekterkennung und -verfolgung besteht in der Detektion von vorher festgelegten Markern. Ein Marker kann ein physisch vorhandenes Bild, Logo oder Barcode sein, der von einer Software erkannt wird. Die Software errechnet in Echtzeit die Pose[4] des Markers und kann dann z. B. ein Objekt auf oder neben den Marker rendern. Solange der Marker im Blick der Kamera ist, kann sich der Benutzer beliebig um den Marker herumbewegen, das Objekt bleibt im Idealfall an seinem vorgesehenen Platz.

Welche Art von Markern sich zur Erkennung besser eignen, ist von der Implementation des Erkennungsalgorithmus abhängig. Manche basieren auf der Erkennung von Kanten und Kontrasten und bevorzugen daher z. B. QR-Codes oder abstrakte Formen (wie etwa in Abbildung 2.1). Andere Algorithmen suchen etwa nach vielen eindeutigen Punkten auf dem Marker und würden daher gleichförmige Grafiken eher schlecht erkennen.

Neben grafischen Markern können auch unauffälligere (wenn auch aufwändigere) Alternativen verwendet werden. Infrarot-basiertes Tracking orientiert sich an für das menschliche Auge unsichtbaren Infrarotstrahlen, die z. B. von Infrarot-LEDs ausgestrahlt werden. CCDs[5] können diese Strahlen sichtbar ma-chen, sodass die Software sie als Marker verwenden kann. Der Vorteil dabei ist, dass auf den ersten Blick kein auffälliger (und evtl. unerwünschter) Marker benutzt werden muss. Allerdings müssen die LEDs mit Strom versorgt werden und sind komplizierter zu handhaben als ein einfacher, ausgedruckter Marker.

---

[4]Verbindung aus Position und Orientierung. Beschreibt Ausrichtung eines Objektes im Raum.
[5]Charged-Coupled Device. Weit verbreiteter Bildsensor-Typ, der in fast allen digitalen Kameras vorhanden ist (vgl. [Blanc01])

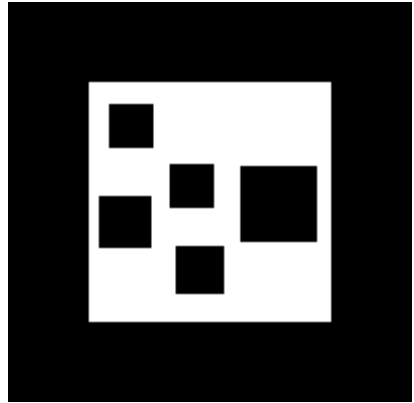

Abbildung 2.1.: Beispiel für einen kontrastbasierten schwarz/weiß Marker

Mit magnetischem Tracking wird in [Lehmann05] eine weitere Möglichkeit vorgestellt. Dabei werden von einem Sender durch drei Spulen (für jede Richtung je eine) Magnetfelder erzeugt, die durch drei Spulen in einem Empfänger detektiert werden können. Ist die Form des Magnetfeldes bekannt, können Rückschlüsse auf die Position des Empfängers geschlossen werden. Vorteile bietet ein magnetisches Trackingsystem z. B. in der Medizin. Da die Magnetfelder nicht auf optischer Basis arbeiten, und Sender- und Empfangsspulen daher keinen direkten „Sichtkontakt" benötigen, sind sie von Verdeckungen weniger betroffen, was im OP-Situs größere Freiheit für die Operateure bedeutet. Außerdem ist auch Tracking über Ultraschallwellen (ebd.), mechanischen Gelenksystemen (ebd.) und durch Inertialkräfte mit Beschleunigungssensoren und Gyroskopen möglich ([FHA98]).

## 2.2.2. Markerloses Tracking

Die andere Möglichkeit besteht in der Erkennung der Eigenschaften eines Objektes. Statt eines Markers nutzt man auffällige und einzigartige Charakteristiken (Kanten, Ecken), die als Anhaltspunkt für die Pose eines Objektes dienen können. Dazu gibt es mehrere Ansätze.

Versucht man die Struktur eines Objektes zu erkennen, müssen die gesuchten Merkmale für eine einfachere Detektion hervorgehoben werden. Soll also ein Modell auf der Basis von Kanten verfolgt werden, bietet sich ein Kantendetektor wie der Canny-Detektor[6] an. Will man einzigartige Punkte verfolgen, können diese mit einem Feature-Detektor wie FAST[7] [Rosten05] oder SIFT[8] [Lowe99] erkannt werden. In beiden Fällen müssen die erkannten Merkmale zwischen zwei Kameraframes verfolgt werden, damit aus diesen die Differenz und die neue Pose des Objektes errechnet werden kann.

Eine andere Möglichkeit ist an das aus der Robotik kommende SLAM (*Simultaneous Localization and Mapping*)-Verfahren angelehnt, bei dem eine Karte des zu verfolgenden Gebietes erstellt wird. Bei PTAM [KM07] (*Parallel Tracking and Mapping*), werden zuerst mit dem FAST-Detektor Bildmerkmale gesucht, auf deren Basis dann eine Karte des betrachteten Gebietes (z. B. ein Schreibtisch) erstellt wird. Die Bildmerkmale werden zwischen zwei Kameraframes verfolgt und können dadurch für AR genutzt werden. Die Karte wird dabei nicht ständig neu erstellt, sondern erweitert.

Mit der Erweiterung *PTAMM*[9] kann zwischen mehreren Karten automatisch gewechselt werden. Außerdem besteht die Möglichkeit sie abzuspeichern und zu laden. Beide Methoden sind allerdings relativ rechenintensiv, die Autoren empfehlen einen Zweikern-Prozessor mit ca. 2 GHz und eine OpenGL-beschleunigte Grafikkarte (Vgl. [KleinRm]).

---

[6]Siehe dazu im Glossar: Canny Detektor
[7]Features from Accelerated Segment Test
[8]Scale Invariant Feature Transform
[9]`http://www.robots.ox.ac.uk/~bob/software/index.html`

Einen ähnlichen Ansatz wie PTAM verfolgt Microsoft momentan mit *Kinect-Fusion* [Izadi11]. Dabei werden die Tiefeninformationen einer Kinect[10] zur Erstellung einer Karte genutzt, außerdem sollen aus der erfasste Szene in Echtzeit in 3D-Modelle rekonstruiert werden.

Eine weitere Option ist modellbasiertes Tracking. Dabei wird ein bereits vorhandenes 3D-Modell des zu verfolgenden Objektes mit dem Kamerabild abgeglichen und dabei versucht, die Pose des realen Objekts mit der des Modells zu vergleichen. Ist der Vergleich erfolgreich, können daraus Korrespondenzpunkte erstellt werden, und diese als Anhaltspunkt für Poseveränderungen genommen werden.

## 2.3. Virtuelle Produktentstehung und digitales Mock-Up

Virtuelle Produktentstehung bezeichnet den computergestützten Analyse, Entwurf und Simulation von Produkten im Rahmen ihres Entwicklungsprozesses. Dazu zählen z. B. CA-Systeme (CAD, CAE, . . . ), die die Mitarbeiter beim Erstellen und Planen unterstützen, oder virtuelle Produktmodelle, die bei der strukturellen bzw. funktionalen Analyse eingesetzt werden (siehe [Schaich01] und [DGF11]).

Der Begriff Digital Mock-Up (DMU) stammt aus der virtuellen Produktentstehung und wird in [KFG07_b] als *„eine virtuelle Attrappe [. . . ], welche die Produktstruktur und die Produktgeometrie repräsentiert"* definiert. Ein DMU ist also kein hochdetailliertes Modell des Produktes, wie es in CAD-Anwendungen entworfen wird, sondern lediglich eine daraus erzeugte geometrische Repräsentation (Vgl. ebd. ). Es beinhaltet außerdem keine Funktion (es könnte z. B. kein

---

[10]Eine ursprünglich für die Xbox 360 Spielekonsole entwickelte Kamera, die Gesten erkennen kann und über einen Tiefensensor verfügt.

Stromfluss analysiert werden), sondern dient nur zur Überprüfung von Form und Oberflächeneigenschaften.

## 2.4. Interaktion

Ziel dieser Arbeit ist die Interaktion mit einem 3D-Modell auf einem mobilen Gerät. Doch was bedeutet Interaktion in diesem Zusammenhang, und resultiert aus Interaktion auch Interaktivität?

Goertz beschreibt Interaktion in [Goertz04] im Kontext der Informatik als: „*Im Verständnis der Informatik bezeichnet ‚Interaktion‘ immer das Verhältnis von Mensch und Maschine, nicht aber die Kommunikation zwischen zwei Menschen mittels einer Maschine.*" Nach diesem Zitat kann man den Begriff „Interaktion" also durchaus auf den Inhalt dieser Arbeit anwenden, schließlich ist die geplante Anwendung für die Kommunikation eines Anwenders mit einem mobilen Gerät gedacht.

Interaktivität hingegen, die nach [LB04] „*die Chance eines einfachen und kontinuierlichen Rollentausches zwischen den Sendern und Empfängern von Informationen*" ermöglicht, ist nach dieser Definition nicht Teil dieser Arbeit. Der Empfänger (hier das Mobilgerät) reagiert nur auf die Eingaben des Senders. Der im Zitat angesprochene Rollentausch findet nicht statt, da der Empfänger außer über eine visuelle Rückmeldung nicht mit dem Sender kommuniziert.

Interaktion im Zusammenhang mit virtueller Realität wird in [KFG07_c] wie folgt beschrieben:

> „*Unter dem Begriff Interaktion sind alle Mechanismen zu verstehen, die die direkte Einflussnahme des Benutzers auf die virtuelle Welt ermöglichen. Dazu gehören sowohl Änderungen des Blickwinkels oder der Position des Benutzers als auch das Auslösen von Prozessen.*"

Das Zitat beschreibt den Begriff in Zusammenhang mit dieser Arbeit sehr treffend. Der Anwender soll über den Touchscreen die Position und Orientierung des Modells verändern. Die Änderung des Blickwinkels erfolgt dann entweder über die grade erwähnte Änderung des Objektes, die Änderung der Position des Markers oder durch den Benutzer selbst, wenn er sich um den Marker herumbewegt. Darüber hinaus soll der Anwender mit dem Modell interagieren können, indem er einzelne Teile des Modells antippt und genauer betrachtet („*direkte Einflussnahme*", ebd. ).

## 2.5. Die Android Plattform

Android ist ein von Google und der Open Handset Alliance[11] entwickeltes Betriebssystem und eine Software-Plattform für mobile Kleingeräte wie Smartphones und Tablet-PCs.

Android ist *open-source*[12], d. h. der Quelltext der Plattform kann von jedem eingesehen, heruntergeladen und verändert werden. Ziel der Offenlegung der Plattform ist es, die Bedürfnisse von Mobilfunkbetreibern, Anwendern und Entwicklern nicht von einer einzigen Firma abhängig zu machen und die Anpassung und Fehlerbehebung durch alle Nutzergruppen möglich zu machen (Vgl. [AOS]).

Android basiert auf dem Linux Kernel 2.6 und besteht aus mehreren Grundanwendungen (wie Browser, Adressbuch und E-Mail Client), einem Anwendungs-Framework, Bibliotheken und einer Laufzeitumgebung (Vgl. [Andr_j]).

Über das Android SDK können in Java Anwendungen für das Betriebssystem geschrieben werden. Das Anwendungs-Framework ist die eigentliche API

---

[11]`http://www.openhandsetalliance.com/` und im Glossar: Open Handset Alliance
[12]Android steht zum Großteil unter der Apache Software License 2.0, Ausnahmen und Details siehe `http://source.android.com/source/licenses.html`

und umfasst den Zugriff auf die Klassen und Methoden, die zum Programmieren für Android nötig sind. Android stellt dem Programmierer verschiedene Fremdbibliotheken wie OpenGL ES (3D-Grafiken) oder SQLite (relationale Datenbank) zur Verfügung, wobei der Zugriff darauf ebenfalls über das Anwendungs-Framework erfolgt. Android-Anwendungen werden in einer virtuellen Maschine (*Dalvik Virtual Machine*, ähnlich der Java VM) ausgeführt, wobei jeder Prozess in seiner eigenen VM läuft (Vgl. ebd.).

## 2.5.1. Aufbau einer Android-Anwendung

Eine Android-Anwendung besteht aus mindestens einer *Activity*. Eine Activity ist ein einzelner Programmteil (meist ein Fenster oder Bildschirm), mit dem der Benutzer interagieren kann [Andr_b]. Ein Adressbuch könnte z. B. aus je einer Activity zum Hinzufügen, Bearbeiten und Versenden einzelner Kontakte und zur Anzeige aller Kontakte bestehen. Dabei kann immer nur eine Activity gleichzeitig aktiv sein. Wechselt man zwischen ihnen, wird die Vorherige pausiert.

Activities können untereinander mit *Intents* aufgerufen werden. Ein Intent ist ein Objekt, das Informationen über eine auszuführende Operation beinhaltet [Andr_c]. Möchte man also bei obigem Adressbuch einen Kontakt anrufen, wählt man dessen Telefonnummer aus. Intern wird dabei eine neue Activity gestartet, die den Intent mit den Informationen `ACTION_DIAL` und `tel:030/123456` enthält. In diesem Fall entscheidet das Betriebssystem, dass die (eingebaute) Activity zum Durchführen eines Anrufes aufgerufen wird.

## 2.5.2. Lebenszyklus von Android Anwendungen

Activities können sich nach [Andr_a] in einem von drei Zuständen befinden:

**Resumed** Die Activity befindet sich im Vordergrund und kann Eingaben vom Benutzer empfangen. Der Zustand wird auch als *Running* bezeichnet

**Paused** Eine andere Activity befindet sich im Vordergrund, verdeckt die andere aber nicht komplett (bspw. ein Optionsmenü)

**Stopped** Die Activity ist von einer Anderen vollständig verdeckt und nicht mehr sichtbar

Android nutzt eine Reihe von Callback-Methoden zur Verwaltung des Lebenszyklus und der Zustände von Activities. Abbildung 2.2 illustriert die Reihenfolge der Methodenaufrufe. Sie werden nicht explizit durch den Programmierer, sondern vom System aufgerufen und müssen in der Klasse der Activity überschrieben[13] werden. Um eine sinnvolle, minimale Anwendung zu erstellen, ist zumindest eine eigene Implementierung von `onCreate()` notwendig.

Die folgende Liste zeigt wann die einzelnen Methoden aufgerufen werden.

**onCreate()** Während die Activity startet

**onStart()** Kurz bevor die Activity sichtbar wird

**onResume()** Die Activity ist sichtbar

**onPause()** Eine andere Activity kommt in den Vordergrund (die Aktuelle wird pausiert)

**onStop()** Die Activity ist nicht mehr sichtbar und wird gestoppt

**onRestart()** Nachdem die Activity gestoppt wurde und wieder in den Vordergrund kommt

**onDestroy()** Die Activity wird zerstört

(Vgl. [Andr_f])

---

[13]Im Sinne von *method overriding* in der objektorientierten Programmierung

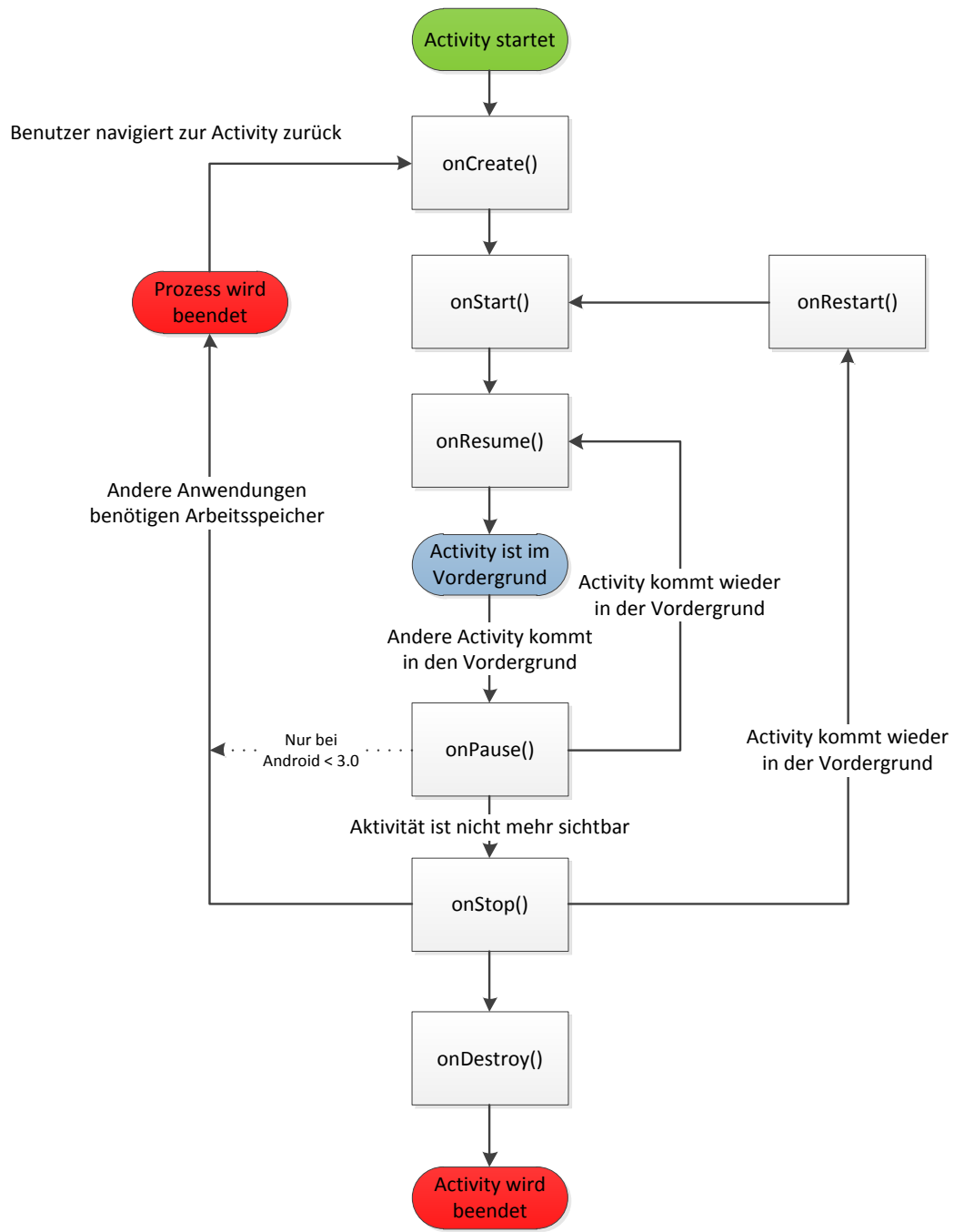

Abbildung 2.2.: Lebenszyklus von Android Activities. Nach [b_Andr].

Dabei ist zu beachten, dass in Android die Activities meist nicht explizit vom Benutzer beendet werden. Ist eine Activity nicht mehr sichtbar, kann sie vom Betriebssystem beendet werden, falls eine andere Anwendung Arbeitsspeicher benötigt. Unterhalb von Android Version 3.0 können auch pausierte Anwendungen jederzeit vom System beendet werden, dies ist in neueren Versionen aber nicht mehr der Fall[14]. Ansonsten bleibt die Activity im Speicher, und ist bei erneutem Aufruf schneller verfügbar.

### 2.5.3. OpenGL ES

OpenGL for Embedded Systems (OpenGL-ES, deutsch: „*OpenGL für einge-bettete Systeme*") ist eine 3D-Grafikschnittstelle speziell für Grafiklösungen im mobilen Sektor und in eingebetteten Systemen (Videospielkonsolen, PDAs). OpenGL ES basiert auf OpenGL, besitzt aber einen geringeren Funktionsum-fang.

Aktuell gibt es zwei relevante Versionen von OpenGL ES, 1.1 und 2.0. Android unterstützt seit Erscheinen v1.1, ab Android 2.2 ist auch v2.0 implementiert [Andr_h].

Version 1.1 funktioniert ähnlich dem ursprünglichen OpenGL (bis Version 3.0) mit einer Fixed Function Pipeline (FFP, deutsch: „*Leitung mit festen Funktio-nen*") [Andr_g]. Die Vertices[15] durchlaufen diese Pipeline bis zur endgültigen Darstellung auf dem Bildschirm. Eine vereinfachte schematische Darstellung der FFP findet sich in Abbildung 2.3.

In der FFP wird die Darstellung über feste Funktionen („fixed functions", Teil der OpenGL API) realisiert, die programmatisch aufgerufen werden und die Vertices begrenzt manipulieren können. Über diese Funktionen kann z. B. auf die Beleuchtung, die auf jeden Vertex angewendet wird, Einfluss genommen werden. Nachdem die Vertices zusammengesetzt und gerastert wurden, sind sie im Prinzip Pixel mit zusätzlichen Farbinformationen und werden *fragments* genannt. Auf die fragments kann dann wieder mit festen Funktionen Einfluss genommen werden, bevor sie nach u. a. Tiefenberechnungen auf dem Bildschirm dargestellt werden.

---

[14]Original: „*Starting with Honeycomb, an application is not in the killable state until its onStop() has returned*" [AAPI_a]

[15]Ein Vertex beschreibt einen Punkt im Raum und kann neben der Position auch Farbin-formationen etc. beinhalten.

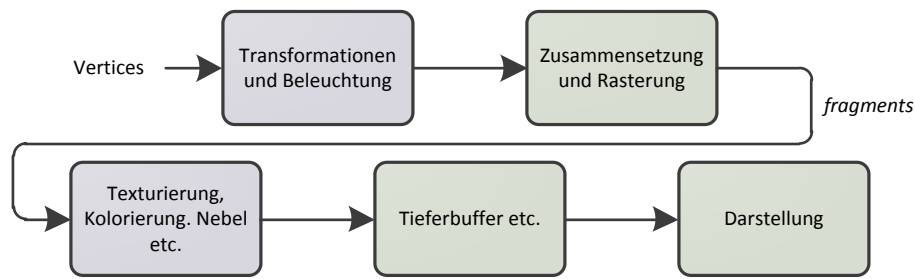

Abbildung 2.3.: Fixed Function Pipeline in OpenGL ES 1.x (vereinfachte Darstellung). Nach [b_KG_b] und [b_RBW_a].

Ab OpenGL ES 2.0 gibt es die FFP nicht mehr, stattdessen wird eine programmierbare Pipeline genutzt[16], mit der kleine Programme (*Shader*) direkt auf der GPU[17] ausgeführt werden. Shader werden bei OpenGL in der GL Shading Language (GLSL) geschrieben, und ersetzen die festen Funktionen der FFP. Wie in Abbildung 2.4 dargestellt, ersetzen *Vertex-Shader* die Transformationen und Beleuchtungseinstellungen der FFP. Nach der Rasterung können die fragments mit einem *Fragment-Shader* modifiziert werden, in dem z. B. Bump-Mapping implementiert werden kann.

Allerdings stellt die Verwendung von Shadern mit GLSL eine erhebliche Umstellung dar, weil selbst für einfache Beleuchtungssituationen ein Shader geschrieben werden muss. OpenGL ES 2.0 ist daher nicht rückwärtskompatibel zu v1.x.

---

[16]Siehe KG_a.
[17]Graphics Processing Unit, deutsch: „*Grafikprozessor*"

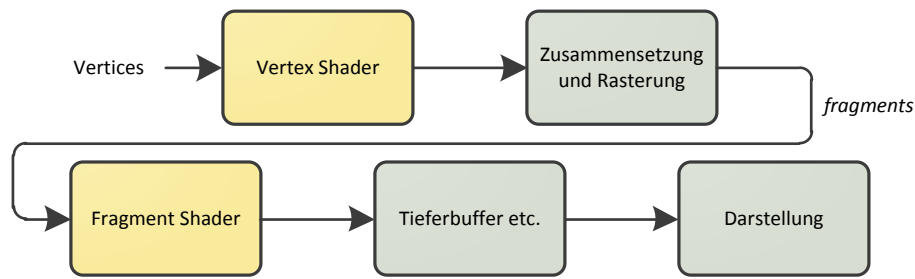

Abbildung 2.4.: Programmierbare Pipeline in OpenGL ES 2.0 (vereinfachte Darstellung). Nach [b_KG_a] und [b_RBW_b].

## 2.5.4. Android Native Development Kit (NDK)

Android unterstützt durch das Native Development Kit (NDK) die Einbindung von C/C++ („nativem Code") in die Java-Umgebung. Das NDK wiederum nutzt das Java Native Interface (JNI), um aus Java heraus nativen Code aufzurufen.

Die Nutzung von C/C++ in Android-Anwendungen kann in bestimmten Fällen Geschwindigkeitsvorteile bringen, insbesondere bei „*CPU-intensiven Operationen die wenig Speicher allokieren, wie Signalverarbeitung, physikalische Simulationen usw.*"[18] [Andr_k].

Neben Geschwindigkeitsgründen spricht auch die Wiederverwendbarkeit von vorhandenem C/C++ Code, der entweder nur aufwändig oder mangels entsprechender Funktionen gar nicht auf Java portiert werden könnte, für die Verwendung des NDKs.

---

[18]Original: „*Typical good candidates for the NDK are self-contained, CPU-intensive operations that don't allocate much memory, such as signal processing, physics simulation, and so on.*"

Eine Reihe von CV-Bibliotheken wie OpenCV[19] oder FastCV[20] nutzen das NDK, um die Einbindung von Algorithmen zur Bildverarbeitung in nativem Code zu ermöglichen und so (als Teil der oben genannten Signalverarbeitung) zu beschleunigen.

Um in nativem Code Daten für die Verwendung durch den aufrufenden Java-Code herzustellen, sind einige besondere Datentypen erforderlich, da C/C++ nicht direkt mit Java-Objekten umgehen kann. JNI bietet dafür Datentypen mit dem Präfix *j* an, z. B. `jint` für den Java int-Typ oder `jfloatArray` für ein Java Array aus float-Zahlen. Nur diese Datentypen können JNI-Methoden übergeben und auch wieder an Java zurückgegeben werden.

### 2.5.5. Interaktionsmöglichkeiten

Die Möglichkeiten, mit denen der Benutzer mit einer Anwendung oder dem Gerät interagieren kann, sind von der verwendeten Hardware auf der Android läuft abhängig.

Im Android Compatibility Definition Document (*CDD*, siehe [GoCD]), das die Richtlinien für Android-Geräte festschreibt, wird für jedes Android 2.3-kompatible Gerät ein berührungsempfindlicher Bildschirm (*Touchscreen*) vorausgesetzt. Der Touchscreen sollte zudem mehrere Eingaben gleichzeitig registrieren können (*Multitouch*). Außerdem sollten die Geräte einen Beschleunigungssensor (Accelerometer), Kompass, GPS-Empfänger und Orientierungssensor (Gyroskop) haben.

Mit diesen Geräten und Sensoren können mehrere Interaktionsmöglichkeiten realisiert werden. Über den Touchscreen kann der Benutzer mit einem oder mehreren Fingern Eingaben vornehmen, oder über Gesten z. B. Bilder und

---

[19]`http://opencv.willowgarage.com/wiki/Android`
[20]`https://developer.qualcomm.com/develop/mobile-technologies/computer-visi on-fastcv`

Webseiten skalieren. Mit einem Accelerometer können Orientierungsänderungen des Gerätes gemessen werden, was genutzt werden kann, um Objekte zu drehen oder zu kippen. Mit einem zusätzlichen Gyroskop sind genaue Positionsbestimmungen im Raum möglich. Kompass und GPS-Empfänger können zur Standortbestimmung genutzt werden, was ortsabhängige Dienste ermöglicht; wie der schon unter 2.1 erwähnte Wikitude AR Browser.

Der Zugriff auf die Sensoren erfolgt über die Android API. Dafür stehen u. a. die `SensorManager`-Klasse für den Zugriff auf die unterschiedlichen Sensoren des Gerätes und die `SensorEvent`-Klasse, die die gemessenen Werte eines Sensors zurückliefert, zur Verfügung.

## 2.6. Zusammenfassung

In der Industrie und Medizin kann AR in Kombination mit vorhandenen Methoden einen Mehrwert in Form von zusätzlichen Informationen schaffen, die kontextsensitiv für den Anwender zur Verfügung stehen.

Die Android-Plattform für mobile Geräte sieht verschiedene Möglichkeiten zur Interaktion mit dem Benutzer vor, die Entwickler ausnutzen können, um Anwendungen im AR-Bereich zu erstellen. Durch die Unterstützung von OpenGL ES ist auch die Darstellung von dreidimensionalen Objekten auf dem Gerät möglich, die Teil einer AR-Anwendung sein können.

Android ermöglicht mit dem NDK die Einbindung von nativem C/C++ Code in Anwendungen, um hardwarenahe Programmierung und die Übernahme von vorhandenem Code zu ermöglichen.

Im nächsten Kapitel werden einige bereits verfügbare AR-Lösungen auf der Android-Plattform und verwandte Arbeiten vorgestellt.

# 3. Stand der Technik

Dieses Kapitel gibt einen Überblick über einige verfügbare Augmented Reality Toolkits für die Android Plattform, und stellt kurz ihre Funktionsweise und Möglichkeiten zur Objektverfolgung vor.

## 3.1. Qualcomm Augmented Reality SDK (QCAR)

Das amerikanische Telekommunikationsunternehmen Qualcomm bietet Entwicklern das QCAR-SDK für die Android- und iOS-Plattformen zur Erstellung von Augmented Reality Anwendungen an. Das SDK ist kostenlos und erlaubt in der derzeitigen Lizenzfassung[21] das Entwickeln und Veröffentlichen von kostenlosen und kommerziellen AR Anwendungen. Die Quellen des SDKs liegen nicht offen (*closed source*), daher ist auch die genaue Funktionsweise des Marker-Erkennungsalgorithmus nicht bekannt. Qualcomm gibt lediglich an, dass die Erkennung auf „natural features" (deutsch: *„natürliche Merkmale"*) basiert (Vgl. [Qcomm_a]).

Als Marker werden von der Software möglichst „unordentliche", d. h. Bilder mit hohen Kontrasten, wenig leeren Flächen und vielen eindeutigen Bildmerkmalen bevorzugt. Die Erzeugung der Marker erfolgt über ein Webinterface, wobei die gewünschten Bilder auf die Server von Qualcomm hochgeladen und analysiert werden. Die Analyse gibt Aufschluss über die Tauglichkeit des Bildes als Marker, wobei diese nur als Empfehlung dient. Genutzt werden können sie in jedem Fall.

---

[21]siehe [Qcomm_b]

Die Bilder werden danach in ein Binärformat umgewandelt und können dann heruntergeladen und in die Projektumgebung kopiert werden. Es können höchstens ca. 50 Marker (Vgl. [Qcomm_a]) gleichzeitig in einem Projekt verwendet werden. Die Marker werden fest in die Anwendung einkompiliert, und können zur Laufzeit nicht verändert werden.

In der Beta Version 1.5.4 des SDKs können die Marker zumindest von einem beliebigen Ort auf dem Gerät geladen werden, und sind so ohne neue Kompilierung der Anwendung austauschbar [Qcomm_c].

QCAR kann die Marker auf unterschiedliche Weise nutzen [Qcomm_d]. Als *Image Target* können bis zu fünf Marker gleichzeitig erkannt und verwendet werden. *Multi Targets* bestehen aus mehreren, fest zueinander gehörenden Markern. So kann z. B. ein Würfel mit sechs unterschiedlichen Seiten als ein einzelner Marker behandelt werden. *Frame Marker* sind ähnlich QR-Codes und haben ein binäres Muster, bestehend aus Quadraten am Rand der Bilder gedruckt. Die Verfolgung diesen Typs Marker ist nicht Merkmal-basiert, d. h. es wird nur das kodierte Muster um das Bild herum zum Tracking genutzt, nicht der Bildinhalt selbst.

Qualcomm liefert zu jedem Markertyp Beispielcode mit dem SDK aus; dieser darf auch als Basis für (kommerzielle) Projekte genutzt werden. Dies ist von Vorteil, da viel Code für die Initialisierung, Markererkennung und Kontrolle des Programmflusses notwendig ist. Die Bibliothek nutzt teilweise das NDK und JNI (siehe Abschnitt 2.5.4) um auf interne Funktionen des SDKs zuzugreifen.

## 3.2. ARToolKit-basierende SDKs

ARToolKit ging 1999 aus einem AR Konferenz-System hervor [KB99] und wurde ursprünglich für normale Computer entwickelt (nicht für mobile Geräte). Für nicht-kommerzielle Verwendung steht das Projekt unter der GPL, bei kommerzieller Verwendung ist eine Lizenzierung von ARToolWorks[22] nötig. Mit AndAR[23] und NyARToolkit[24] existieren Portierungen auf die Android-Plattform. Allerdings ist keines der Projekte auf einem aktuellen Stand und damit auch nicht an neuere Version von Android angepasst. Seit 2007 ist keine neue Version von ARToolKit erschienen, die letzte Projektaktivität von AndAR stammt aus dem Jahre 2010[25]. Von NyARToolkit ist zumindest eine Version für das 2011 veröffentlichte Android 3.0 verfügbar.

ARToolKit und seine Derivate nutzen markerbasiertes Tracking, wobei einfache Marker mit schwarz/weißen Mustern (ähnlich Abbildung 2.1) verwendet werden. Um Marker in ein Projekt zu integrieren wird eine Webcam benötigt, die über ein ARToolKit beliegendes Programm einen Marker scannt und dessen Tracking-Informationen in einer Datei speichert. Diese Informationen werden von der erstellten Anwendung eingelesen und können dann als Marker verwendet werden.

Problematisch ist die spärliche Dokumentation zu AndAR und NyARToolkit. Zu AndAR sind auf der Projektseite mehrere Beispielanwendungen mit Quellcode verfügbar, eine ausführliche Dokumentation wie bei QCAR existiert nicht. Da ARToolKit ursprünglich aus Japan stammt, und NyARToolkit ebenfalls dort entwickelt wird, sind nur maschinenübersetzte Versionen der API-Dokumentation aus dem Japanischen existent, was das Entwickeln verkompliziert.

## 3.3. Sonstige Software und ähnliche Arbeiten

In diesem Abschnitt werden der Vollständigkeit halber noch zwei weitere Anwendungen näher betrachtet. Neben dem AR-SDK der Firma metaio, was

---

[22]http://www.artoolworks.com/
[23]http://code.google.com/p/andar/
[24]http://sourceforge.jp/projects/nyartoolkit-and/
[25]http://code.google.com/p/andar/source/list

aufgrund hoher Lizenzkosten nur kurz erläutert wird, wird außerdem eine An-
wendung von Mercedes Benz, die einen ähnlichen Zweck, wie die in dieser
Arbeit geplanten Implementierung verfolgt, näher beleuchtet.

### 3.3.1. Metaio Mobile SDK

Das Mobile SDK für Android und iOS der in München ansässigen metaio
GmbH gibt es in einer kostenlosen Basisversion und einer rund €20.000 teu-
ren Pro-Version. Soll eine Anwendung ohne Wasserzeichen von metaio erstellt
werden, müssen zusätzlich rund €5000 in der Basisversion bzw. rund €10.000
in der Pro-Version bezahlt werden.

Beide Versionen enthalten eine 3D-Engine mit Shader-Unterstützung und ver-
schiedene Tracking-Möglichkeiten auf optischer Basis (mit QR-Code ähnliche
Markern, Bildern, markerloses 2D-Tracking) oder mit Sensoren (GPS, Ac-
celerometer, Kompass). Mit der Pro-Version ist zusätzlich markerloses 3D-
Tracking möglich.

### 3.3.2. Mercedes-Benz C-Klasse-App

Mercedes-Benz hat zu Demonstrationszwecken eine AR-Anwendung für iOS-
und Android-Geräte entwickelt[26], mit der sich an einem Modell einer C-Klasse
Limousine Anbauteile austauschen lassen können (siehe Abbildung 3.1). Ähn-
lich der Anwendung, die im Rahmen dieser Arbeit entwickelt werden soll, wird
das 3D-Modell des Autos auf einen Marker projiziert. Über vier Schaltflächen
lassen sich dann vier verschiedene zusätzliche Teile, wie z. B. ein Heckspoiler,
direkt am Auto darstellen und wieder entfernen. Darüber hinaus gibt es in der
Anwendung aber keine zusätzlichen Interaktionsmöglichkeiten oder Funktio-
nen.

---

[26]Die Anwendung lässt sich im Android Marketplace kostenlos herunterladen:
`https://market.android.com/details?id=com.MercedesBenzAccessoriesGmbH.`
`arcklasse&hl=de`

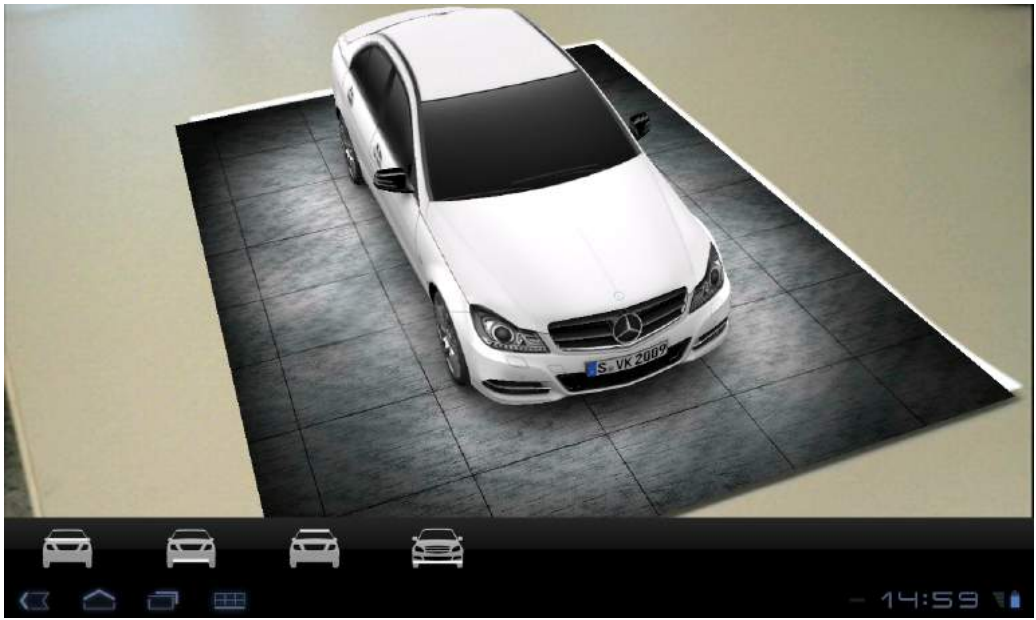

Abbildung 3.1.: Screenshot der Mercedes-Benz C-Klasse-App Quelle: [b_EB]

## 3.4. Zusammenfassung

Das Kapitel hat einige der am Markt erhältlichen AR-Lösungen betrachtet, und ihre Funktionalität hinsichtlich der Markererkennung und -verfolgung beschrieben. Außerdem wurde eine schon vorhandene Anwendung vorgestellt, die simple Interaktion mit AR auf einem mobilen Gerät verbindet.

In Kapitel 4 werden die Anforderungen an die zu entwerfende Anwendung ermittelt und denkbare Szenarien zur Verwendung beschrieben.

# 4. Anforderungsanalyse

In diesem Kapitel werden einige mögliche Anwendungsfälle für die zu erstellende Applikation vorgestellt. Danach werden daraus die benötigten Funktionen für die Anwendung abgeleitet, sowie die Anforderungen hinsichtlich der Software-Qualität analysiert und festgelegt.

## 4.1. Kontext der Anwendung

Mit der Benutzung von bisherige AR- oder VR-Möglichkeiten, wie CAVEs oder HMDs ist jedes Mal ein Aufwand verbunden. In einer CAVE müssen die Projektoren gestartet und Rechner hochgefahren werden, dazu wird ein kompletter Raum belegt. HMDs müssen passend zum Anwender justiert werden, sind ergonomisch gesehen relativ unbequem und schränken die Bewegungsfreiheit entweder durch eine feste Verbindung zum Rechner oder Tragen eines mobilen Computers ein. Der Ansatz dieser Arbeit stellt eine unkomplizierte und ergonomisch bequeme Alternative dazu dar. Ein Tablet-PC ist leicht, kann problemlos zu anderen Projektmitarbeitern getragen werden und erfüllt nicht nur den Zweck der AR-Darstellung.

## 4.2. Mögliche Szenarien

Im Folgenden werden einige mögliche Szenarien vorgestellt, in denen das Projekt Anwendung finden könnte.

## 4.2.1. Kundenberatung und -präsentationen

Autohersteller bieten gegen Aufpreis oft eine Vielzahl an unterschiedlichen Ausstattungsmöglichkeiten an. Diese reichen von anderen Lackierungen über stärkere Motoren bis hin zu verschiedenen Farbstilen der Inneneinrichtung. Aus Kosten- und Platzgründen können Autohäuser nicht jedes Modell in jeder Farbe mit allen Ausstattungsmerkmalen vorrätig haben. Bisher kann dazu auf Computermodelle zurückgegriffen, bei denen man einfach die gewünschten Extras einstellen, und dann dem Kunden präsentieren kann (Vertriebskonfigurator). Einen realistischen Größeneindruck kann man so aber nicht gewinnen.

Mittels AR und markerbasiertem Tracking ließe sich ein Automodell in realistischen Proportionen darstellen. Das Auto würde entsprechend groß auf dem Marker dargestellt werden, über die Anwendung auf einem Tablet-PC oder Smartphone soll der Kunde das Modell von allen Seiten betrachten können. Da vorher in die Anwendung alle relevanten Farbvarianten und optischen Ausstattungsmerkmale einprogrammiert wurden, könnten sie über Schaltflächen schnell verändert und direkt am Modell angezeigt werden. Alternativ könnten auch einige Bereiche des Automodells per Berührung anwählbar sein (z. B. die Felgen oder das Dach), wobei dem Kunden dann über ein Textfenster Informationen angezeigt werden, oder er aus einem erscheinenden Menü unterschiedliche Felgenvarianten oder ein zusätzliches Schiebedach auswählen könnte.

Der Vorteil gegenüber einer Präsentation an einem stationären PC liegt in der intuitiven Bedienbarkeit. Wie bei einem physisch vorhandenen Auto könnte der Kunde mit dem Tablet-PC um den Marker (und somit um das Modell) herumgehen, und es von der Nähe und aus der Entfernung betrachten.

Darüber hinaus könnte der Kunde die Anwendung (sofern er ein geeignetes mobiles Gerät besitzt) und den Marker problemlos mit nach Hause nehmen und sich das Modell samt Ausstattung vor dem Kauf noch einmal in vertrauter

Umgebung im Detail ansehen – unabhängig von der Größe der Wohnung, da das 3D-Modell skaliert werden kann.

## 4.2.2. Bauraumprüfungen

Bei der Entwicklung von Produkten finden stetig Anpassungen am aktuellen Prototypen statt. Da DMUs primär Form und Oberfläche veranschaulichen, können sie zum Beispiel für die Einpassung und Sichtprüfung[27] von Bauteilen eingesetzt werden. Wird festgestellt, dass das überprüfte Teil nicht wie vorgesehen passt, kann mit der geplanten Implementierung das Modell zumindest passend skaliert werden.

Mittels einer erweiterten Version könnten auch noch direkt auf dem Gerät kleinere Anpassungen vorgenommen werden, etwa das Vergrößern eines Bohrloches. In diesem Fall liegt der Vorteil darin, dass die Änderungen nicht endgültig sind. An einem physischen Modell kann man einmal entferntes Material nicht wieder anbringen.

Die Daimler AG nutzt AR in diesem Zusammenhang, um die Befestigung und den optischen Eindruck von kleineren Bauteilen zu kontrollieren (Vgl. [Specht11], Folie 15).

## 4.2.3. Entwurfsprüfungen

Ähnlich dem oben genannten Fall, soll die Anwendung auch bei Entwurfsprüfungen (engl. *Design Review*) die Ingenieure unterstützen. Bei der Überarbeitung einer vorhandenen Komponente (z. B. einer Lichtmaschine) soll das AR Modell direkt über das echte Bauteil gelegt werden können, um mit dem bereits verbesserten DMU die Änderungen im direkten Vergleich zu sehen. Da das AR-Modell möglichst exakt auszurichten ist, muss die Interaktionen auf einzelne Achsen beschränkt werden können, da eine freie Positionierung im Raum den Vorgang nur erschweren würde.

---

[27]In diesem Fall mit Hilfsmitteln: Einer Kamera und einem AR-Modell

Damit das Anzeigen des Modells nicht nur auf den Marker beschränkt wird, soll das Modell vom Tracking entkoppelt werden und auch ohne Marker betrachtet und verändert werden können. Das ist dann hilfreich, wenn der AR-Aspekt nebensächlich ist, und z. B. nur eine Veränderung mit einem Kollegen oder bei einer Teambesprechung besprochen werden soll.

## 4.3. Identifizierte Anwendungsfälle

Zusammenfassend lassen sich aus oben genannten Einsatzmöglichkeiten die Anwendungsfälle in Abbildung 4.1 abbilden.

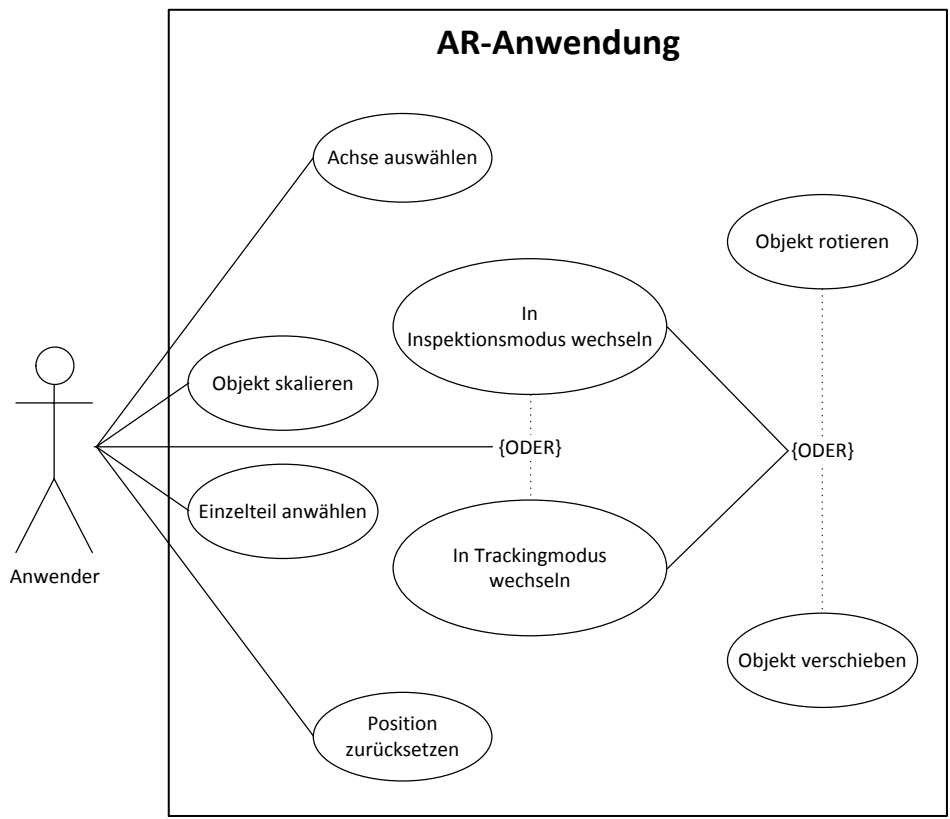

Abbildung 4.1.: Identifizierte Anwendungsfälle

- Der Benutzer soll das Objekt skalieren können

- Der Benutzer soll die Position des Objektes zurücksetzen können

- Es soll eine Achse (X, Y, Z) ausgewählt werden können, auf die die Transformationen beschränkt werden

- Zur einfacheren Bedienbarkeit soll entweder nur Rotiert oder nur Verschoben werden können

- Es soll mit einzelnen Teilen des Modells interagiert werden können

- Um das Objekt nicht rein auf den Marker zu beschränken, soll es aus dem AR-Kontext („Trackingmodus") herausgelöst, und unabhängig von ihm betrachtet und transformiert werden können („Inspektionsmodus")

## 4.4. Anforderungen an das Android-Gerät

Damit der Benutzer die Anwendung möglichst komfortabel benutzen kann, sollte das Mobilgerät eine Reihe von Merkmalen bzw. integrierten Funktionen besitzen. Da die Bedienung ausschließlich über den Bildschirm erfolgen soll, wird ein Touchscreen benötigt. Um die Skalierung des Modells ebenfalls über den Touchscreen realisieren zu können, ist es notwendig, dass er Multitouch[28] fähig ist.

Die zur Darstellung benötigten 3D-Modelle sind faktisch reduzierte CAD-Modelle und somit für Mobilgeräte vergleichsweise komplex, daher sollte die Prozessor- und Grafikleistung des Gerätes möglichst hoch sein. Zwar sollte die Anwendung auch auf schwächer ausgestatteten Geräten laufen, allerdings sind dann Einschränkungen bei der Darstellungsgeschwindigkeit zu erwarten. Da das QCAR SDK mindestens Android 2.1, einen ARMv6-Prozessor und eine FPU[29] vorraussetzt[30], gilt dies ebenfalls für die geplante Implementierung.

---

[28]Der Bildschirm muss mehrere Eingaben (durch mehrere Finger) gleichzeitig verarbeiten können.

Das QCAR SDK nutzt OpenGL ES 2.0, deswegen ist ein entsprechend kompatibles Gerät notwendig. QCAR kann zwar auch mit OpenGL ES 1.x verwendet werden, allerdings werden dafür mehrere Änderungen im Quellcode notwendig, und es müsste eine separate Version kompiliert werden.

## 4.5. Qualitätsmerkmale

Ausgehend von den Anwendungsfällen und Einsatzszenarien werden Qualitätsanforderungen an die Anwendung gestellt, die in Tabelle 4.1 vermerkt sind. Nach ISO/IEC 9126 [DSD] werden dabei die Kriterien *Funktionalität, Zuverlässigkeit, Benutzbarkeit, Effizienz, Änderbarkeit* und *Übertragbarkeit* herangezogen. Auf die besonders relevanten, bzw. explizit nicht relevanten Merkmale dieser Kriterien, wird im Folgenden eingegangen.

Da in der Anwendung die Benutzerinteraktion eine wichtige Rolle spielt, sind insbesondere die *Bedienbarkeit* und *Erlernbarkeit* entscheidend. Die Anwendung soll für den Benutzer ohne große Einarbeitungszeit nutzbar sein und möglichst problemlos und intuitiv bedienbar sein. Außerdem soll das System unkompliziert auf sich ändernde Anforderungen (primär der Austausch von Modellen und Markern) einstellbar sein. Die *Modifizierbarkeit* ist daher als sehr wichtig einzustufen, außerdem darf die *Stabilität* dadurch nicht beeinträchtigt werden.

Die *Ordnungsmäßigkeit* ist hingegen eher nebensächlich. Da keine persönlichen Daten verwendet werden, und die Anwendung mehr oder weniger ein reines Informationssystem ist, steht dieser Punkt nicht im Fokus. Sollte später

---

[29]Floating Point Unit, deutsch: „*Gleitkommaeinheit*". Ein dedizierter Prozessor zur Durchführung von Gleitkommazahl-Berechnungen

[30]`https://ar.qualcomm.at/qdevnet/support_file/download/Android/supported de vices android/Supported_Devices_Android.pdf`

aber ein Rechte-System integriert werden, das verschiedene Benutzergruppen (z. B. „Kunde" und „Ingenieur") implementiert, ist eine sichere Trennung von sensiblen und öffentlichen Daten allerdings zu beachten. Durch die Android-Plattform ergeben sich außerdem Kriterien der *Übertragbarkeit.* Da die Installation von Android-Anwendungen entweder durch den Marketplace oder über USB-Kabel durch den Entwickler erfolgen muss, können keine besonderen Anforderungen an die *Installierbarkeit* gestellt werden. Bei Android läuft zudem jede Anwendung unabhängig[31] von der anderen, außerdem sind sie in einer einzelnen Datei gespeichert wobei das Betriebssystem deren Organisation übernimmt. Daher sind die *Koexistenz* und *Austauschbarkeit* nicht relevant.

---

[31]Jeder Prozess läuft in einer eigenen VM (Vgl. [Andr_e])

| Merkmal | sehr wichtig | wichtig | normal | nicht relevant |
|---|---|---|---|---|
| **Funktionalität** | | | | |
| Angemessenheit | | x | | |
| Interoperabilität | | | x | |
| Ordnungsmäßigkeit | | | | x |
| Richtigkeit | | x | | |
| Sicherheit | | | x | |
| **Zuverlässigkeit** | | | | |
| Fehlertoleranz | | x | | |
| Konformität | | x | | |
| Reife | | x | | |
| Wiederherstellbarkeit | | | x | |
| **Benutzbarkeit** | | | | |
| Attraktivität | | | x | |
| Bedienbarkeit | x | | | |
| Erlernbarkeit | x | | | |
| Konformität | | | x | |
| Verständlichkeit | | x | | |
| **Effizienz** | | | | |
| Konformität | | x | | |
| Verbrauchsverhalten | | | x | |
| Zeitverhalten | x | | | |
| **Änderbarkeit** | | | | |
| Analysierbarkeit | | x | | |
| Konformität | | | x | |
| Modifizierbarkeit | x | | | |
| Stabilität | x | | | |
| Testbarkeit | | | x | |
| **Übertragbarkeit** | | | | |
| Anpassbarkeit | | x | | |
| Austauschbarkeit | | | | x |
| Installierbarkeit | | | | x |
| Koesxistenz | | x | | x |
| Konformität | | | x | |

Tabelle 4.1.: Qualitätsmerkmale der Anwendung

# 5. Entwurf

Nachdem in Kapitel 4 die Anforderungen an das System analysiert und beurteilt wurden, wird nun ein Entwurf der Anwendung beschrieben.

## 5.1. Wahl der Android Version

Die Anwendung wird für Android 3.x entwickelt, da in dieser Version insbesondere Verbesserungen für Geräte mit großen Bildschirmen, wie z. B. Tablet-PCs, eingeführt wurden. Für diese Arbeit interessant ist hier vor allem die neue *Actionbar*, die Schaltflächen am oberen Bildschirmrand (ähnlich Tabs in aktuellen Webbrowsern) bereitstellt. Hintergrund dafür ist die einheitliche Platzierung von Schaltflächen in verschiedenen Anwendungen (siehe [Andr_d]).

Da aktuell die Zahl der Geräte mit Android Version 2.2 und 2.3.3 stark überwiegen, und der Marktanteil der 3.x Versionen bei wenigen Prozent liegt[32], wird aus Kompatibilitätsgründen aber auch eine Alternative (mittels des herkömmliche Optionsmenüs) zur Actionbar eingebaut. Da mittelfristig nur noch Geräte mit neueren Android Versionen (aktuell ist Version 4) auf den Markt kommen, soll diese Anpassung später wieder entfernt werden können.

---

[32]Siehe die Android Statistiken von Januar 2012: [Andr_i]

## 5.2. 3D-Engine

Die Darstellung eines 3D-Modell ist ein wichtiger Bestandteil dieser Arbeit. Da die Anwendung modifizierbar sein soll, ist es wichtig, dass ein oder mehrere (beliebige) 3D-Modelle ohne großen Aufwand geladen und ausgetauscht werden können. Üblicherweise können CAD-Modelle in gebräuchliche Formate wie 3DS[33] oder OBJ[34] exportiert werden[35], allerdings bietet OpenGL von Haus aus keine Möglichkeit an, solche Formate zu laden. Hier bietet sich die Verwendung einer 3D-Engine an, die u. a. das Laden und Darstellen solcher Dateiformate ermöglicht und als eine Art Abstraktionsschicht zwischen OpenGL und dem Programmierer fungiert. Dadurch wird z. B. die Kapselung von 3D-Modellen in Objekte (sodass Transformationen über Methodenaufrufe auf das Objekt angewendet werden können), oder das einfache Hinzufügen von Lichtquellen oder primitiven Objekten[36] ermöglicht.

## 5.3. Benötigte Komponenten

In diesem Abschnitt werden die unterschiedlichen Komponenten beschrieben, aus denen die Anwendung bestehen soll. Aufgrund der Architektur von Android ergibt sich schon eine Besonderheit: Die Benutzeroberfläche ist vom OpenGL-Renderer entkoppelt, da er standardmäßig in einem eigenen Thread läuft [AAPI_b].

### 5.3.1. Benutzeroberfläche

Aus dem Anwendungsfalldiagramm aus Abbildung 4.1 lässt sich bereits eine Komponente ableiten; die Benutzeroberfläche. Es werden daher benötigt:

---

[33]Autodesk 3ds Max Format
[34]Wavefront Object Format
[35]Beispielsweise aus Rhino3D [McN] und AutoCAD [Adesk], oder mittelbar über Programme wie AccuTrans 3D [MMP].
[36]engl. *primitives*. Grundlegende 3D-Objekte wie Würfel oder Ebenen

- Je eine Schaltfläche für die Translation, die Rotation und das Zurücksetzen des Modells

- Einen Schalter um zwischen Inspektions- und Trackingmodus zu wechseln

- Ein Dropdown-Menü zur Auswahl der Achsen

Die Skalierung des Modells soll über eine Geste auf den Multitouch-Bildschirm erfolgen und benötigt daher keine eigene Schaltfläche. Damit Translation und Rotation konsistent (mit einem Finger) bedient werden können, wird auf die Implementierung einer Multitouch-Geste zur Rotation verzichtet. Ferner bietet Android schon einen besonderen Listener[37] für eine 2-Finger-Skalierung an (siehe Abschnitt 6.3.1), nicht aber für eine 2-Finger-Rotation.

Abbildung 5.1 zeigt eine Grafik der Benutzeroberfläche mit den oben identifizierten Objekten.

Unten befindet sich die standardmäßige Systemleiste von Android, mit den Schaltflächen (von links nach rechts) zum Beenden und Minimieren der Anwendung, dem Anwendungsumschalter und dem fakultativen Optionsmenü.

Oben, in der Actionbar, befinden sich die Schaltflächen zum Transformieren, Moduswechsel, Zurücksetzen und ein Dropdown-Menü zur Auswahl der Achsen. Hier zeigt sich ein weiterer Vorteil der Actionbar – die Schaltflächen können direkt berührt werden, ohne dass erst vorher ein Optionsmenü geöffnet werden muss. Nur die verschiedenen Achsen wurden der Übersicht halber in einem Untermenü platziert.

Die Benutzeroberfläche ist die Haupt-Activity der Anwendung. Da es in der geplanten Anwendung keine weiteren Einsatzmöglichkeiten für Activities gibt, bleibt dies auch die Einzige. Das hat zur Folge, dass sich der Code zur Verarbeitung der Eingaben in der gleichen Klasse befindet, wie der Steuerungscode des AR-Teils. Eine Kapselung in separate Klassen ist nicht ohne weiteres möglich, da beide Komponenten Zugriff auf die Haupt-Activity benötigen und im Falle einer Trennung zusätzlicher Code erforderlich wäre.

---

[37]deutsch: „*Zuhörer*" . Ein Objekt das bei einem bestimmten Ereignis automatisch benachichtigt wird und darauf reagieren kann.

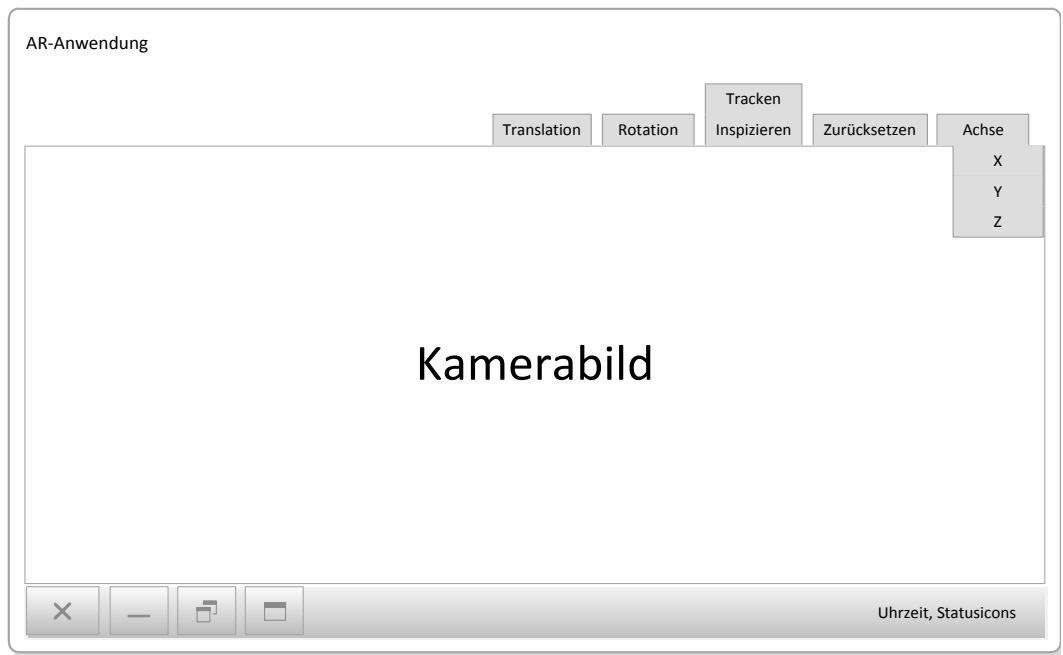

Abbildung 5.1.: Prototyp der Benutzeroberfläche

In Hinblick auf Objektorientierung ist diese „Vermischung" nicht ideal, allerdings beeinträchtigen oder beeinflussen sich beide Komponenten nicht. Lediglich die Übersicht in der Klasse leidet etwas.

### 5.3.2. OpenGL-Renderer

In der zweiten Komponente, dem Renderer, findet die Darstellung des Modells statt. Die Eingaben auf dem Touchscreen bei Android sind Teil der Benutzeroberfläche, und die erfassten Werte werden über eine Rückruffunktion an die Klasse, die die Oberfläche implementiert, gesendet. Ebenso werden die gewählten Menüpunkte aus der Actionbar verarbeitet.

Da die Oberfläche wie schon erwähnt in einem anderen Thread läuft, besteht die Möglichkeit, dass *race conditions*[38] zwischen Benutzeroberfläche und Renderer auftreten. Die Android API bietet dazu eine Lösung in Form einer Queue an.

Um die Eingabedaten passend zum aktuellen Modus zu verarbeiten, sind eine Reihe von Abfragen notwendig. Konkret muss beachtet werden:

- Hat überhaupt eine Berührung stattgefunden? Falls nicht, muss auch der aktive Modus nicht berücksichtigt werden.

- Ist der Translations- oder Rotationsmodus aktiv? Die Eingabedaten müssen dann entweder als Rotation oder Translation auf das Modell angewendet werden.

- Ist der Inspektionsmodus aktiv? Dann muss keine neue Pose berechnet werden, da das Modell nicht auf den Marker projiziert werden soll.

- Hat eine Interaktion stattgefunden? Falls ja muss das berührte Teil des Modells bestimmt, und dessen Informationen angezeigt werden.

Die Struktur der Renderer-Klasse ist in Abbildung 5.2 als Programmablaufplan dargestellt.

---

[38]Eine Situation, in der zwei oder mehr Prozesse gleichzeitig auf gemeinsam genutzte Daten zugreifen, ohne sich gegenseitig darüber zu informieren. Dadurch kann es zu Inkonsistenz der Daten kommen.

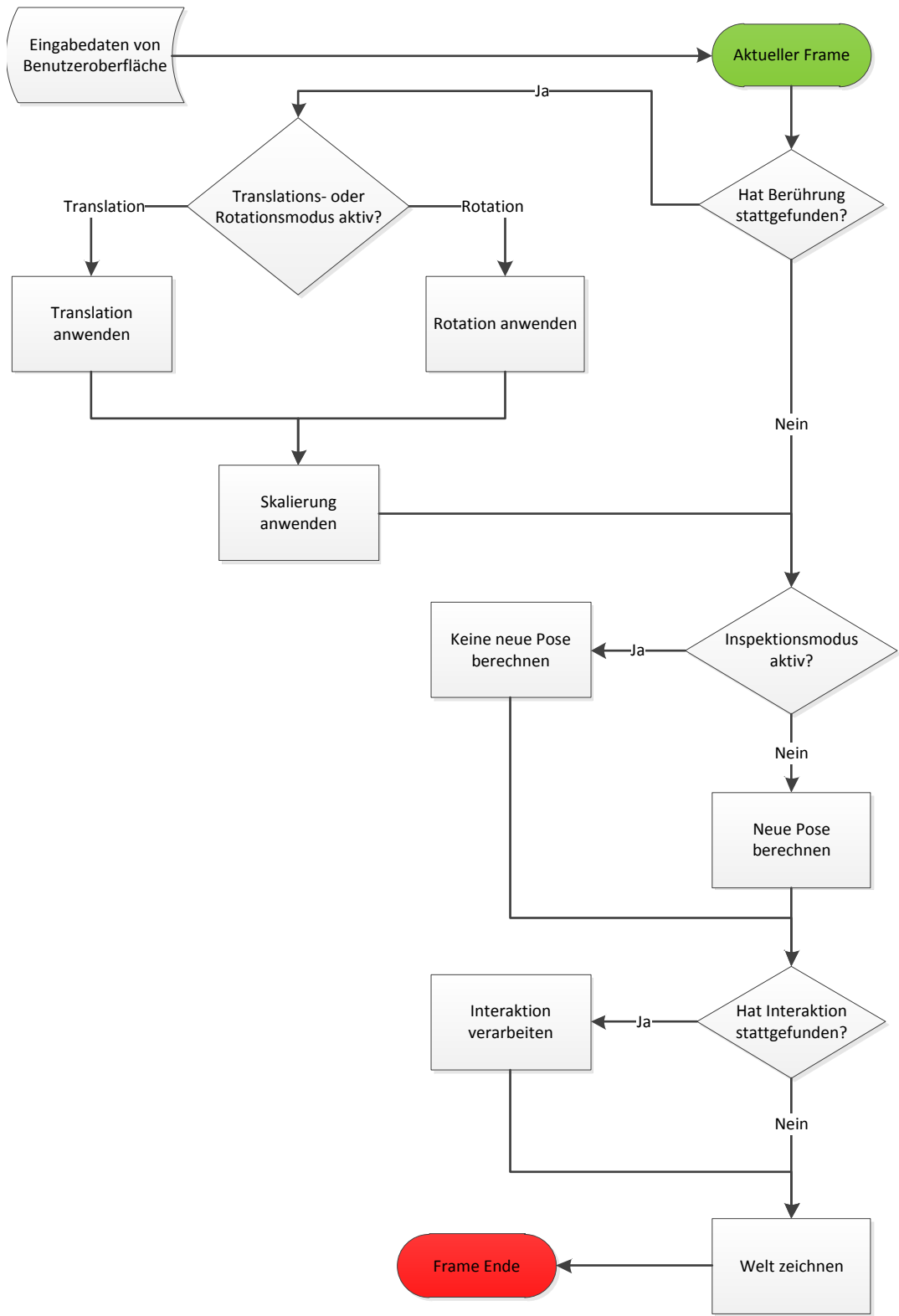

Abbildung 5.2.: Ablaufplan der Datenverarbeitung in der Renderer-Klasse

### 5.3.3. Markerberechnung

Zur Verfolgung des Markers wird das QCAR-SDK von Qualcomm verwendet. Das SDK hat eine Reihe von Vorzügen gegenüber anderen Produkten. Ein wichtiger Aspekt ist das stabile und robuste Tracking, das auch noch bei spitzen Winkeln und größeren Entfernungen (bis etwa. 2 Meter mit einem DIN A4 großen Marker) noch zuverlässig die Pose bestimmen kann. Darüber hinaus wird ausführlicher Beispielcode bereitgestellt, der das Erstellen des relativ komplexen Aufbaus der Anwendung vorweg nimmt und daher auch als Grundlage für diese Arbeit dient. Zuletzt spielen neben der kostenlosen Verfügbarkeit auch die aktuell noch fortlaufende Weiterentwicklung des SDKs, sowie die native Implementierung für Android eine Rolle.

Die Steuerung der AR-Komponenten (Initialisierung der Kamera, Marker und des Trackings; sowie deren Deinitialisierung) befindet wie unter 5.3.1 beschrieben in der Haupt-Activity.

Damit die Anwendung funktioniert, muss der Marker zwingend erkannt werden. Allerdings muss er nicht vollständig von der Kamera erfasst werden, da QCAR der Marker auch erkennt, wenn nur ein Teil von ihm für die Kamera sichtbar ist.

#### 5.3.3.1. Orientierung des Markers

Neben der Pose des Markers ist auch dessen reine Orientierung interessant. Um dem Benutzer eine möglichst einheitliche Bedienung bei der Verschiebung des Modells zu ermöglichen, muss die Richtung, in die verschoben wird, unter Umständen invertiert werden. Am folgenden Beispiel eines Autos als 3D-Modell wird dargelegt, warum das notwendig ist. So soll z. B. eine Verschiebung auf der x-Achse immer einer Verschiebung auf der Längsachse[39] entsprechen.

---

[39]Im Falle eines Autos eine gedachte Achse vom Motorraum bis zum Kofferraum.

Abbildung 5.3 zeigt das gewollte Verhalten bei Verschiebung des Modells. Der Benutzer führt auf dem Touchscreen eine Rechtsbewegung aus, um das Auto nach rechts zu verschieben. Der Renderer würde also z. B. die Anweisung „Verschiebe das Objekt um 50 Einheiten auf der x-Achse" bekommen.

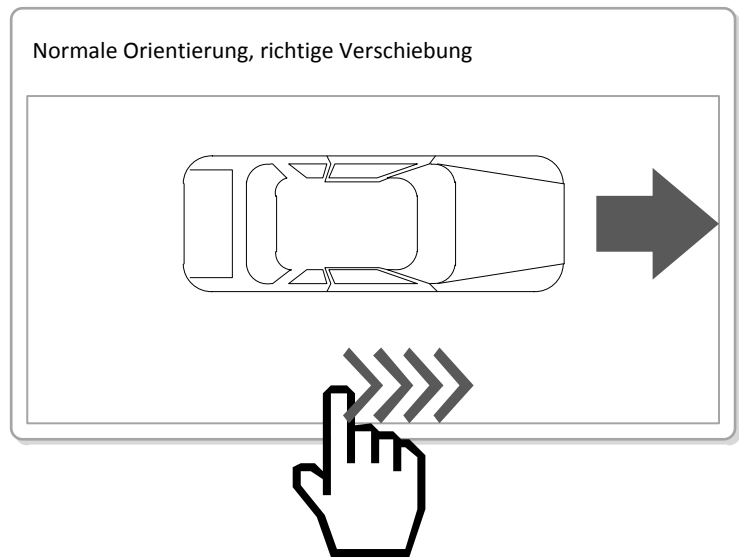

Abbildung 5.3.: Korrekte Verschiebung des Modells bei normaler Orientierung

Wenn der Benutzer sich nun auf die andere Seite des Autos bewegt, oder der Marker um 180° gedreht wird, muss sichergestellt werden, dass sich das Modell bei einer Rechtsbewegung auf dem Touchscreen auch nach rechts verschiebt. Die Anweisung „Verschiebe das Objekt um 50 Einheiten auf der x-Achse" würde hier das Modell aber in die falsche Richtung verschieben, da der Renderer nicht wissen kann, in welcher Relation wir uns zum Auto befinden und das Modell in dessen Objektkoordinatensystem nach rechts verschiebt.

In Abbildung 5.4 wird das ungewollte Verhalten dargestellt. Um dies zu umgehen, wird bei der Verschiebung des Modells auch die Rotation des Markers in Verhältnis zum Benutzer berücksichtigt. In diesem Fall muss das Auto um -50 Einheiten auf der x-Achse verschoben werden.

Der Einfachheit halber sollen nur vier verschiedene Fälle berücksichtigt werden. Die „normale" Orientierung, also die Ausgangsposition und um 90°, 180° und 270° rotierte Positionen. Dazwischenliegende Werte werden gleichmäßig den vier Fällen zugeordnet.

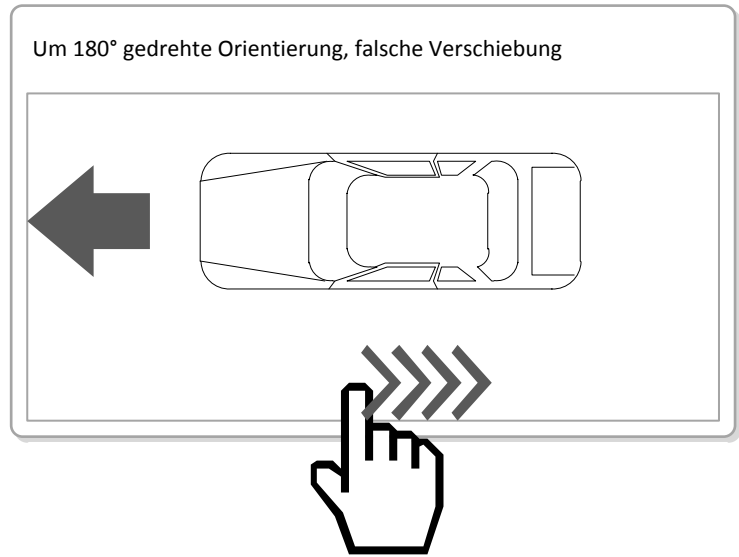

Abbildung 5.4.: Falsche Verschiebung des Modells bei gedrehter Orientierung

# 6. Implementierung

Nachdem in den vorangegangenen Kapiteln die Anforderungen an die Anwendung aufgestellt wurden und daraus ein Entwurf des Systems entstanden ist, wird nun die Implementierung als Android-Applikation beschrieben. Als Objekt wird ein 3D-Modell eines Autos genutzt, das dem Fraunhofer IPK zu Forschungszwecken zur Verfügung steht und hier in reduzierter Form (nur die äußere Form, ohne Technik und Innenraum) zum Einsatz kommt. Das Modell liegt im OBJ-Format vor und ist intern aus mehreren Gruppen (wie Unterboden, Motorhaube oder Fenster) zusammengesetzt.

## 6.1. Entwicklungsumgebung

Das Android SDK liefert alle nötigen Werkzeuge, um Android-Anwendungen zu kompilieren und auf das Gerät zu laden. Zusätzlich ist ein ADT (*Android Development Tools*)-Plugin für die Entwicklungsumgebung Eclipse[40] verfügbar, das den Compiler, Debugger und Tools wie *Logcat* (zeigt fortlaufend u. a. Debug-Ausgaben und Statusmeldungen an) oder GUI-Editoren zur Oberflächengestaltung direkt in Eclipse integriert, und die Verwendung der sonst nutzbaren Android Kommandozeilen-Tools obsolet macht.

Weil Zugriff auf das NDK nötig ist, müssen außerdem Eclipse CDT (C/C++ Development Tools) und das NDK an sich installiert werden.

---

[40]`http://www.eclipse.org/`

Da Android auf dem Linux-Kernel basiert, nutzt das NDK zum Kompilieren der C-Quelltexte eine Reihe von Tools (wie *make* oder *awk*) aus der Linux-Welt, um daraus Binärpakete für die ARM-Architektur (die primäre CPU-Architektur der Android Plattform [ARM]) zu erstellen. Um diese auf einem Windows-PC nutzen zu können, ist Cygwin[41] notwendig, das die benötigten Linux-Tools nach Windows portiert.

Entwickelt wird die Anwendung auf einem *LG V900 Optimus Pad* Tablet-PC mit Android 3.1.

## 6.2. 3D-Engine - jPCT-AE

Zur Darstellung des 3D-Modells soll, wie in Abschnitt 5.2 beschrieben, eine 3D-Engine genutzt werden, um möglichst einfach das Modell laden und austauschen zu können. Da darüber hinaus keine besonderen Anforderungen an die 3D-Darstellung gestellt sind, wird die Engine *jPCT-AE*[42] genutzt. jPCT-AE kann verschiedene 3D-Dateiformate wie z. B. OBJ oder 3DS laden, wurde speziell für Android portiert und optimiert [Foerster_b] und kann einfache 3D-Umgebungen erstellen, ohne dass direkt mit OpenGL gearbeitet werden muss.

Eine insbesondere für den mobilen Bereich interessante Möglichkeit von jPCT ist das Laden von serialisierten Modellen[43]. Der Vorteil hierbei liegt in der schnelleren Ladezeit gegenüber nicht-serialisierten Modellen, die sich auf mobilen Geräten mit relativ beschränkten Ressourcen bemerkbar macht.

Das Modell wird in seinen Einzelteilen in ein `Object3D` Array geladen, auf das dann Transformationen angewendet werden können. Darüber wird auch die in Abschnitt 4.3 geplante Interaktion mit einzelnen Teilen realisiert.

---

[41]`http://www.cygwin.com/`
[42]`http://www.jpct.net/jpct-ae/`, der Name steht für *Java Perspective Correct Texture-mapping - Android Edition*, vgl. [Foerster_a] und [Foerster_b]
[43]Siehe dazu im Glossar: Serialisierung

## 6.3. Verarbeitung der Eingabedaten

In diesem Abschnitt wird der Datenfluss von einer Berührung auf dem Touchscreen bis zur Verarbeitung dieser Daten durch den Renderer beschrieben.

### 6.3.1. Benutzeroberfläche

Berührt der Benutzer den Touchscreen, werden einige Daten in einem Objekt über eine Rückruf-Methode an die Activity gesendet. Aus diesem Objekt lassen sich unter anderem der Typ der Berührung (Finger berührt Bildschirm, Bewegung über den Bildschirm, Finger berührt nicht mehr den Bildschirm etc.) und die aktuelle Position der Berührung abrufen.

Wie in 5.3.2 erwähnt müssen die Eingabedaten des Touchscreens threadsicher an den Renderer übergeben werden. Dazu bietet Android die `queueEvent()`-Methode der `GLSurfaceView`-Klasse an. Ein SurfaceView ist eine Art Oberfläche, auf der unterschiedliche Objekte angezeigt werden. Im Unterschied zu einem normalen `View` kann aber ein separater Thread genutzt werden (hier der OpenGL Thread), der auf diese Oberfläche rendert. Die `queueEvent()`-Methode platziert beliebige Daten in eine Queue, die der Renderer dann verarbeitet, sobald er einen neuen Frame zeichnet.

Listing 6.1 zeigt einen Ausschnitt aus der Methode `onTouchEvent()`, die immer dann aufgerufen wird, wenn eine Berührung oder Bewegung auf dem Touchscreen registriert wurde.

Wenn eine Bewegung auf dem Touchscreen erkannt wurde, sich also ein Finger über den Bildschirm bewegt, ohne dass er zwischendurch der Kontakt mit der Oberfläche unterbricht, werden die aktuellen Positionskoordinaten im switch/case-Konstrukt abgerufen. Dort werden die Koordinaten und ein Signal für den Renderer, dass eine Berührung stattfindet, in der Queue platziert.

Die Skalierung des Modell sollte, wie in Abschnitt 5.3.1 geplant, über eine Geste auf dem Bildschirm erfolgen. Diese wurde über einen `ScaleGestureDetector` implementiert. Die Skalierung erfolgt über eine „Pinch"-Geste (von engl. *to pinch*, deutsch: „*kneifen*"), bei der der Benutzer zwei Finger (normalerweise Daumen und Zeigefinger) auf den Bildschirm setzt und beide dann entweder voneinander weg (vergrößern) oder aufeinander zu (verkleinern) bewegt. Der Detektor liefert einen beliebig eingrenzbaren Wert (um zu große oder zu kleine Werte zu vermeiden) zurück, wobei der Wert 1,0 die Originalgröße darstellt und sich in der vorliegenden Implementation zwischen 0,3 und 5,0 bewegen kann.

Um dem Benutzer bei Erstgebrauch die Bedienung zu erleichtern, wird abhängig von der Position des Markers (siehe 6.3.4) eine Textur mit einem Koordinatensystem eingeblendet, die die Richtung der drei Translations- bzw. Rotationsachsen visualisiert. Über das Optionsmenü lässt sich das Koordinatensystem bei Bedarf aus- und wieder einblenden.

Listing 6.1: Ausschnitt aus der Verarbeitung der Touchscreen-Eingaben

```java
public boolean onTouchEvent(final MotionEvent motionEvent) {

   //Auf Berührungsart reagieren
   switch (motionEvent.getAction()) {

   //Im Fall einer Bewegung über den Touchscreen
   case MotionEvent.ACTION_MOVE:

      //Kommuniziere Daten mit Renderer
      mGlView.queueEvent(new Runnable(){

       public void run() {
         renderer.touchEventFired = true;
         renderer.receivedTouchX = (int) motionEvent.getX();
         renderer.receivedTouchY = (int) motionEvent.getY();
       }
      });

    break;
    //...
    }
  return true;
}
```

### 6.3.1.1. Verarbeitung der Schaltflächen

In Abbildung 5.1 wurde ein Prototyp der Benutzeroberfläche entworfen. Die resultierende Implementierung in Android wird in Abbildung 6.1 dargestellt. Oben befindet sich die Actionbar mit den Schaltflächen, in der rechten oberen Ecke das Menü zur Auswahl der Achsen. Es wurde zusätzlich eine Funktion zur Aktivierung des Blitzlichtes eingebaut, um auch bei schlechten Lichtverhältnissen den Marker erkennen zu können.

Bei der Betätigung eines Schalters wird eine Rückrufmethode aufgerufen, der die ID der Schaltfläche als Argument übergeben wird. Anhand dieser ID werden die Schaltfläche identifiziert und dem Renderer die passenden Variablenwerte übergeben. Zum Beispiel wird bei der Auswahl einer Achse eine Variable in der Renderer-Klasse gesetzt, die die aktuell aktive Achse enthält. Bei einer Transformation wird dann abhängig von dieser Variable die Transformation nur auf diese Achse des Modell angewendet.

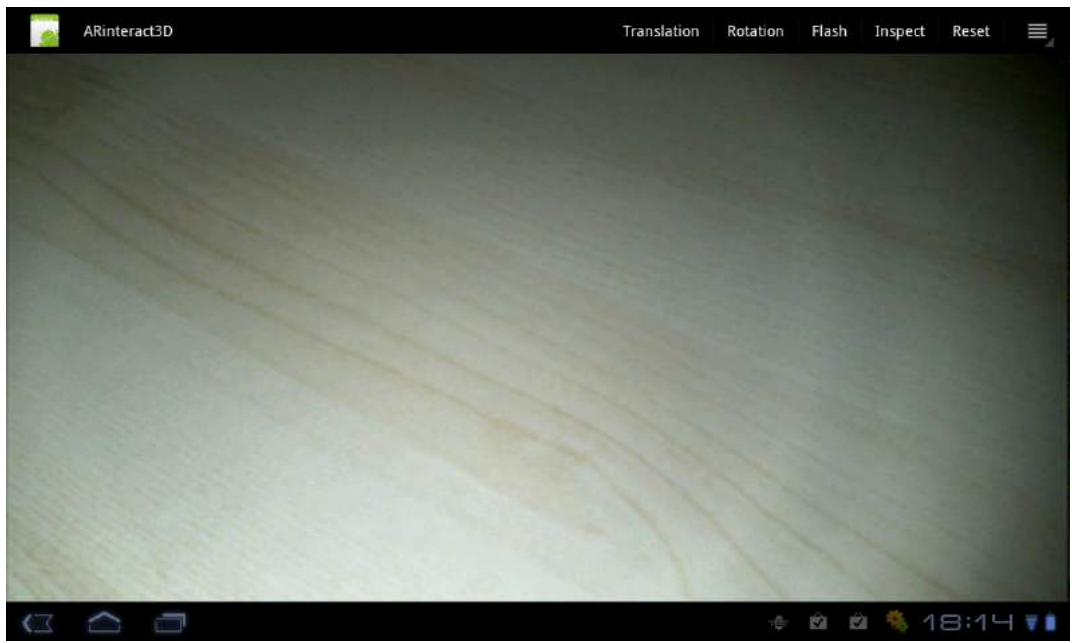

Abbildung 6.1.: Benutzeroberfläche der Anwendung

## 6.3.2. Renderer

In der Renderer-Klasse existiert ebenfalls eine Rückruffunktion, die jedes Mal dann aufgerufen wird, wenn der Renderer einen neuen Frame zeichnet. In der `onDrawFrame()`-Methode ist der Ablauf wie in Abbildung 5.2 implementiert.

Von dem Beispiel aus Listing 6.1 ausgehend, wird dort die Verarbeitung einer Berührung ausgelöst. Ob der Translations- oder Rotationsmodus aktiv ist, wird über eine Variable ermittelt, die von Haupt-Activity gesetzt wird. Standardmäßig ist der Translationsmodus aktiv. Der Wert, um den das Modell verschoben wird, errechnet sich aus der Differenz der X- oder Y-Koordinate (abhängig davon, auf welcher Achse verschoben wird) der letzten und aktuell registrierten Berührungsposition.

Da bei der Errechnung der Markerpose (siehe Absatz 6.3.3) keine vorherigen Transformationen mit einbezogen werden, und immer vom Mittelpunkt des Markers ausgegangen wird, müssen die Werte für Rotation und Translation während der Laufzeit gespeichert werden, d. h. es werden alle bisherigen Rotations- und Translationswerte aufsummiert und auf die neue Pose angewendet. Bei der Skalierung ist dies nicht nötig, da der `ScaleGestureDetector` immer einen absoluten Wert (von der Originalgröße ausgehend) zurückgibt.

## 6.3.3. AR-Darstellung

Nachdem die Translations- oder Rotationswerte verarbeitet wurden, ruft der Renderer eine native (C++)-Funktion auf, die die Markererkennung des QCAR SDKs beinhaltet. Ein Ausschnitt aus dieser Funktion mit den Translationsbefehlen befindet sich in Listing 6.2. Dieser wird der Wert für die Skalierung des Modells und ein jeweils drei Elemente großes Array übergeben, die für jede Achse die Translations- bzw. Rotationswerte beinhalten.

Die QCAR-Funktion `getTrackable()` gibt ein Objekt mit Informationen über den gefundenen Marker zurück, aus dem die aktuelle Pose als 4x4 Matrix ausgelesen wird. Die Transformationswerte werden dann nacheinander (in der Reihenfolge Skalierung, Rotation, Translation) direkt auf die Matrix angewendet. Deren Werte werden in ein Java-kompatibles float-Array kopiert, das die transformierte Matrix zur Verwendung durch die 3D-Engine enthält.

Die zurückgegebene Matrix wird dann als Rotationsmatrix des Modells gesetzt. Nach diesem Schritt ist die Verarbeitung der Transformationsdaten beendet. Das Modell wird entsprechend seiner Position auf den Marker gerendert (siehe Abbildung 6.3).

Listing 6.2: Transformation der Marker-Pose (Ausschnitt)

```
1  JNIEXPORT jfloatArray JNICALL
2  Java_de_fraunhofer_ipk_ARi3D_Renderer_transformMatrix(JNIEnv *env, jobject,
       jfloat scaleAmount, jfloatArray rotations, jfloatArray translations) {
3
4    //Neues Array für die transformierte Pose
5    jfloatArray newPose = env->NewFloatArray(16);
6    jfloat* pNewPose = env->GetFloatArrayElements(newPose, 0);
7
8    //Verschiebungsdaten einlesen
9    jfloat* pTranslations = env->GetFloatArrayElements(translations, 0);
10
11   //Neues Matrix-Objekt erstellen und AR-Tracking beginnen
12   QCAR::Matrix44F poseMatrix;
13   QCAR::State state = QCAR::Renderer::getInstance().begin();
14
15   //Aktiven Marker suchen und dessen Pose auslesen
16   for (int tIdx = 0; tIdx < state.getNumActiveTrackables(); tIdx++) {
17     const QCAR::Trackable* trackable = state.getActiveTrackable(tIdx);
18     poseMatrix = QCAR::Tool::convertPose2GLMatrix(trackable->getPose());
19   }
20
21   //Translationen anwenden, danach die Matrix rotieren und skalieren
22   SampleUtils::translatePoseMatrix(pTranslations[0], 0.0, 0.0, &
         modelViewMatrix.data[0]);
23   SampleUtils::translatePoseMatrix(0.0, pTranslations[2], 0.0, &
         modelViewMatrix.data[0]);
24   SampleUtils::translatePoseMatrix(0.0, 0.0, pTranslations[1], &
         modelViewMatrix.data[0]);
25
26   //Zusätzliche Rotation, um das QCAR-Koordinatensystem dem 3D-
         Koordinatensystem entspricht
27   SampleUtils::rotatePoseMatrix(90.0, 1.0, 0.0, 0.0, &modelViewMatrix.data
         [0]);
28
29
```

```
30   //Matrix in ein float Array kopieren
31   for (int i = 0; i < 16; i++) {
32     pNewPose[i] = poseMatrix.data[i];
33   }
34
35   //Resourcen freigeben
36   env->ReleaseFloatArrayElements(newPose, pNewPose, 0);
37   env->ReleaseFloatArrayElements(translations, pTranslations, 0);
38
39   //AR-Tracking beenden und transformierte Pose zurückgeben
40   QCAR::Renderer::getInstance().end();
41   return newPose;
42 }
```

QCAR verwendet, im Gegensatz zu dem Koordinatensystem des verwendeten 3D-Modells, ein um -90° um die X-Achse rotiertes Koordinatensystem für den Marker (siehe Abbildung 6.2). Daher wird das 3D-Modell nach einer Rotation des Benutzers noch zusätzlich um 90° um die X-Achse rotiert, außerdem werden die Translationen auf der Y- und Z-Achse vertauscht.

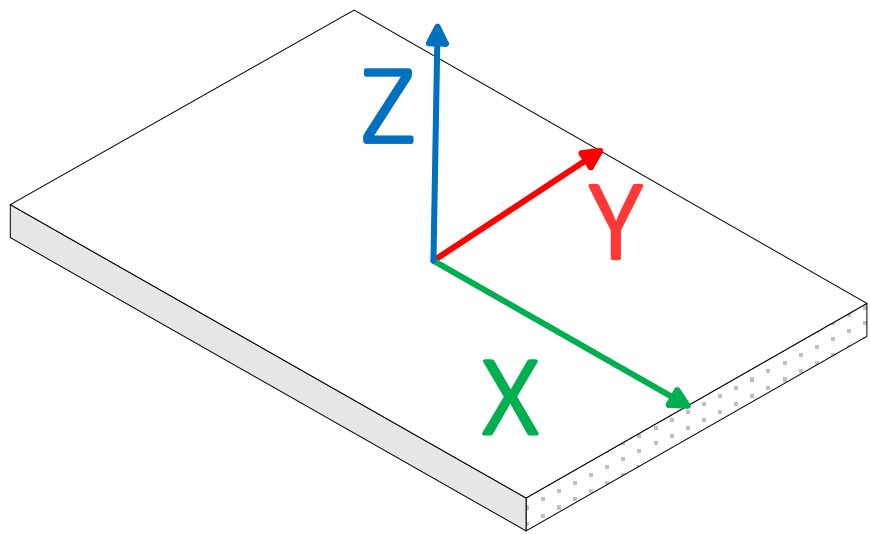

Abbildung 6.2.: Koordinatensystem des Markers

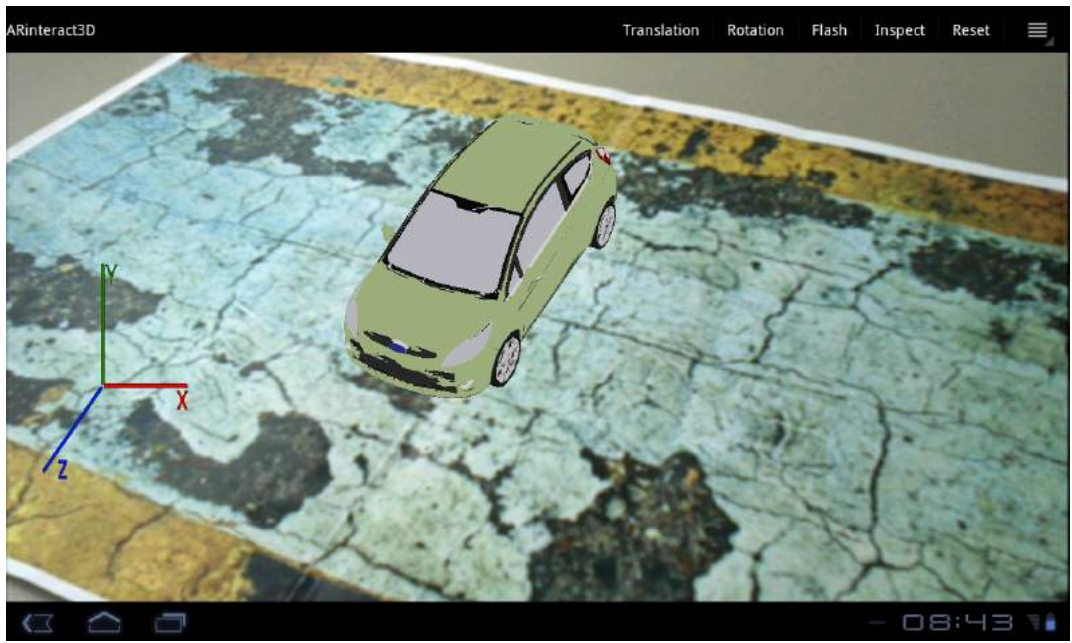

Abbildung 6.3.: 3D Modell wird auf den Marker gerendert

### 6.3.4. Feststellen der Marker-Orientierung

Um die in Abschnitt 5.3.3.1 erwähnte Bedienungsproblematik hinsichtlich der Orientierung bzw. Rotation des Markers zu lösen, wird eine Methode benutzt, die die Orientierung des Markers von in Form eines Rotationswertes von 0 bis 360° zurückgibt. Dabei befindet sich der Marker bei 0° und 180° ungefähr quer zur Kamera, bei 90° und 270° längs.

Werte, die zwischen den vier eben Genannten liegen, werden ihnen gleichmäßig zugeordnet. Der Marker wird demnach in „Normalposition" (Querformat) angenommen, wenn der berechnete Winkel zwischen 315° und 45° liegt. Eine 90° Rotation (Hochformat) wird z. B. für Winkel größer 45° bis 135° festgelegt.

Das Vorgehen und der Quelltext für diese Methode stammen aus dem AR-Entwickler Forum von Qualcomm [ksiva], im Folgenden wird die Berechnung erklärt.

Um die Orientierung des Markers zu berechnen, wird zunächst die gleiche Posen-Matrix, wie bei der normalen AR-Darstellung benutzt. Die Inverse dieser Matrix A wird transponiert, um so die Position der Kamera im Objektraum $A^{-T}$ zu erhalten (Vgl. [KG_b]).

$$A = \begin{bmatrix} r_1 & r_4 & r_7 & 0 \\ r_2 & r_5 & r_8 & 0 \\ r_3 & r_6 & r_9 & 0 \\ t_1 & t_2 & t_3 & 1 \end{bmatrix}, \quad A^{-T} = \begin{bmatrix} r_1 & r_2 & r_3 & t_1 \\ r_4 & r_5 & r_6 & t_2 \\ r_7 & r_8 & r_9 & t_3 \\ 0 & 0 & 0 & 1 \end{bmatrix}^{-1}$$

Der Positionsvektor $\vec{p} = (t_1, t_2, t_3)$ der Kamera lässt sich dann aus der letzen Spalte von $A^{-T}$ ablesen. $\vec{p}$ wird dann auf die X/Y-Ebene (den Marker) in Richtung der Mitte des Markers projiziert und normalisiert. Daraus resultiert der neue Vektor $\vec{v}_{proj}$.

Um nun den Winkel zu berechnen wird das Skalarprodukt $s$ aus $\vec{v}_{proj}$ und dem „Oben"-Vektor $\vec{y} = (0, 1, 0)$ gebildet. $\arccos(s)$ ergibt den Winkel $w$ zwischen den beiden Vektoren im Bogenmaß, die Umrechnung nach Grad erfolgt durch $w = w\frac{180}{\pi}$. $w$ hat nun einen Wert zwischen 0° und 180° (da der Arkuskosinus nur definierte Ergebnisse zwischen 0 und $\pi$ hat).

Um das Vorzeichen von $w$ zu bestimmen (und $w$ auf einen Winkel zwischen 0° und 360° zu bringen), muss das Kreuzprodukt $\vec{v}_{cross}$ aus $\vec{v}_{proj}$ und $\vec{y} = (0, 1, 0)$ berechnet werden. Ist die z-Komponente von $\vec{v}_{cross}$ positiv, so ist auch $w$ positiv (0° bis 180°); ist sie negativ, ist auch $w$ negativ (-180° bis 0°). In beiden Fällen müssen 180° addiert werden, um letztendlich die Orientierung des Markers zurückgeben zu können.

Abhängig von der Orientierung werden dann Translationen auf der X/Z-Ebene (entspricht der Oberfläche des Markers) gegebenenfalls invertiert. Die Y-Achse ist nicht betroffen, da eine Verschiebung in die Höhe unabhängig von der Orientierung ist.

## 6.4. Interaktion mit einzelnen Modellteilen

Um dem Benutzer die Interaktion mit dem Modell zu ermöglichen, muss bei einer Berührung des Touchscreens überprüft werden, ob sich im Bereich der Berührung ein Teil des Modells befindet. Dazu wird die Kollisionserkennung von jPCT-AE genutzt. Beim Laden des Modells wird jedes Einzelteil (hier insgesamt ca. 80) in ein Feld des für das Objekt benutzten Arrays gespeichert. Da die einzelnen Felder nicht unmittelbar zu einem Bauteil zugeordnet werden können[44], muss durch Ausprobieren festgestellt werden, welches der Objekte im Array welchem Bauteil entspricht.

Um festzulegen, ob ein Teil von der Kollisionsabfrage mit einbezogen wird, wird das Attribut `COLLISION_CHECK_OTHERS` gesetzt, das anderen Objekten ermöglicht, mit Diesem zu kollidieren. Dies ist nur einmal nötig und kann etwa beim Laden des Modells bei Programmstart passieren.

Die Kollisionsprüfung erfolgt am Ende der `onDrawFrame()`-Methode, nach der Anwendung aller Transformationen. Zunächst werden die 2D (Touchscreen)-Koordinaten in einen Richtungsvektor im 3D-Raum (in dem sich das Modell befindet) umgewandelt. Aus diesem Vektor und der Kameraposition[45] berechnet jPCT dann das nächstgelegene Polygon und gibt das zugehörige Objekt und die Distanz zu diesem zurück.

Bei der ersten Verwendung eines Modell ist nun die Zuordnung von Einzelteilen zu Array-Elementen nötig. Die interne ID des gewählten Teils wird dafür bei Berührung ausgegeben, sodass einem Array-Element das entsprechenden Bauteil zugeordnet wird. Der Einfachheit halber werden den Einzelteilen aussagekräftigere Namen gegeben („Motorhaube", „Dach"), die beim Laden des Modells gesetzt werden. Die Zuordnung von Teilen und Array-Elementen ist pro geladenem 3D-Objekt immer gleich, insofern muss dies für jedes Modell nur einmal gemacht werden (wenn das Modell später nicht ausgetauscht wird). Diese Namen werden zusammen mit den anderen Modellinformationen in einer Enumeration abgelegt (siehe Abschnitt 6.4.1).

---

[44]Die Namen der Polygongruppen in der OBJ-Datei sind in der vorliegenden Datei nur durchnummeriert. Abhängig vom genutzten 3D-Modell können die Gruppen aber auch schon eindeutige Namen besitzen, was diese Zuordnung erleichtern würde.

[45]Der virtuellen Kamera im 3D-Raum, nicht der Physischen im Gerät

Nachdem die Bezeichnungen einmal vergeben wurden, wird der Name des berührten Objekts mit den Informationen der Enumeration abgeglichen. Die Anzeige der Informationen erfolgt über eine Android-Textbox, wofür ein Handler genutzt werden muss, da das Anzeigen und Ausblenden der Textbox nur durch den Thread der Benutzeroberfläche möglich ist. Dem Handler wird eine Nachricht geschickt (entweder der Informationstext oder „cancel“, um die Textbox auszublenden), der daraufhin entweder den Text anzeigt, oder ihn ausblendet.

Abbildung 6.4 zeigt ein Beispiel dieser Situation. Der Benutzer hat das Dach des Automodells berührt. Das Dach wurde daraufhin etwas nach oben verschoben, oben links wird der Informationstext angezeigt.

Listing 6.3: Auswählen einzelner Modellteile

```java
public boolean findTouchedObject(int touchX, int touchY) {

    //Berechne Richtungsvektor und finde Objekt
    SimpleVector dir = Interact2D.reproject2D3DWS(camera, frameBuffer,
        touchX, touchY).normalize();
    Object[] collisionResult = world.calcMinDistanceAndObject3D(camera.
        getPosition(), dir, 10000);

    //Objekt gefunden?
    if (collisionResult[0] != COLLISION_NONE) {

        //Einzelteil auswählen und vorher gewähltes Teil zurücksetzen
        pickedObject = (Object3D) collisionResult[1];
        previouslyPickedObject.getTranslationMatrix().setIdentity();

        //Informationen aus Enumeration holen
        String information = CarParts.getInformation(pickedObject.getName());

        //Falls Informationen gefunden wurden
        if(!information.equals("NONE")){

            //Entferne alte Textbox und zeige neue an
            messageHandler.sendMessage("cancel");
            messageHandler.sendMessage(information);
```

```
23        //Verschiebung aus Enumeration holen und anwenden
24        float[] transform = CarParts.getTranslation(pickedObject.getName());
25        pickedObject.translate(transform[0], transform[1], transform[2]);
26      }
27
28      //Um bei der nächsten Auswahl das Teil zurücksetzen zu können
29      previouslyPickedObject = pickedObject;
30    }
31  }
```

## 6.4.1. Kapselung der Modell-Informationen

Zu dem gewählten Einzelteil wird ein beliebiger Informationstext in einer Textbox im oberen linken Bereich des Bildschirms angezeigt, außerdem wird es um einen geringen Wert vom Modell weg verschoben, um es davon abzuheben.

Um die zum gewählten Teil passenden Informationen zu speichern, wurde ein eigener Enumerations-Typ[46] erstellt. Dies hat zwei Vorteile. Zum einen lässt sich ein Teil der spezifischen Informationen (Text, Verschiebung) für ein Modell vom restlichen Quelltext entkoppeln. Dadurch wird die Austauschbarkeit dieser Komponente erhöht. Die Enumeration ist für zukünftige Modelle vom restlichen Programm unabhängig, lediglich in der Interaktionsmethode (Listing 6.3) müsste der verwendete Typenname angepasst werden.

Zum anderen wird damit ein Problem bei der Zuordnung von Informationen zu Bauteilen gelöst. Java unterstützt in Version 6 keine Strings in switch-Statements. Erst in Version 7 ist das möglich, welche mit dem Android SDK allerdings noch nicht nutzbar ist. Da das gewählte Einzelteil über dessen Namen (als String) identifiziert wird, wäre eine Zuordnung über ein switch-Konstrukt nicht möglich gewesen.

Alternativ könnten den Objekten über eine if-Abfrage pro Teil ein numerischer Wert zugewiesen werden, der dann wiederum in einem switch-Statement abgefragt würde. Der Nachteil dabei wäre, dass die doppelten Abfragen (switch und if) nicht zur Lesbarkeit beitragen und weniger wartbar wären, als die Lösung mit einer Enumeration. Im Gegensatz zu Strings können Enum-Konstanten in switch-Strukturen genutzt werden und so als Ersatz genutzt werden.

---

[46]In Java definiert das `enum` Schlüsselwort eine Klasse und wird engl. *enum type* genannt, Vgl. `http://docs.oracle.com/javase/tutorial/java/javaOO/enum.html`

Listing 6.4 zeigt einen Ausschnitt aus der Implementation der Klasse. Die Rückgabe des Verschiebungs-Wertes erfolgt ähnlich wie die der Information, nur über ein float-Array mit 3 Feldern (für X/Y/Z-Verschiebung).

Listing 6.4: Enumerations-Typ für Bauteile

```java
public enum CarParts {

  //Bauteile definieren
  MOTORHAUBE, DACH, NONE;

  //Objekt zu Namen finden
  private static CarParts checkName(String name) {
    try {
      return valueOf(name);
    }
    //NONE zurückgeben, falls kein Objekt gefunden oder Name übergeben wurde
    catch (IllegalArgumentException iae) {
      return NONE;
    }
    catch (NullPointerException npe) {
      return NONE;
    }
  }

  //Anhand des Namen die Information zurückgeben
  public static String getInformation(String partname) {
    String information = "NONE";

    switch (checkName(partname)) {

    case MOTORHAUBE:
      information = "Beliebiger Text";
      break;

    case DACH:
      information = "Beliebiger Text";
      break;
    }

    return information;
  }
```

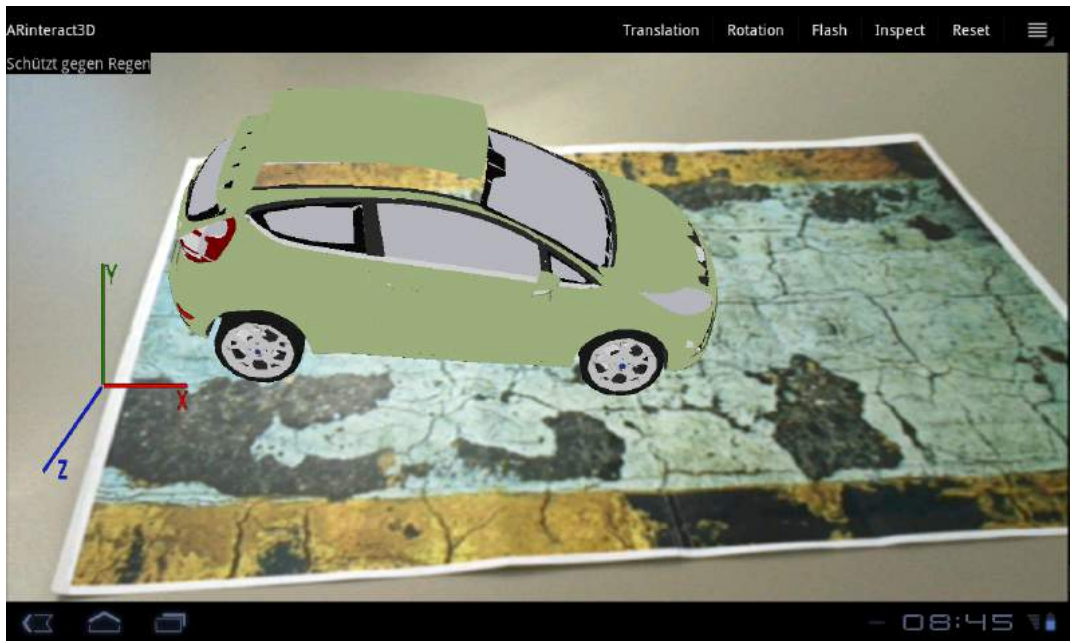

Abbildung 6.4.: Berührtes Modellteil mit Hervorhebung und Informationstext

## 6.4.2. Inspektionsmodus

Damit der Anwender das Modell auch unabhängig von einem Marker betrachten kann, wurde ein Modus implementiert, der das Tracking stoppt und das Modell in seiner aktuellen Position auf dem Bildschirm belässt. Der Anwender kann immer noch mit dem Modell interagieren und es transformieren, muss das Gerät aber nicht auf den Marker zeigen lassen.

Die Funktionsweise ist ähnlich dem Trackingmodus, der in Abschnitt 6.3.3 beschrieben wurde. Im Unterschied dazu wird aber keine neue Pose des QCAR Trackers geholt, sondern die Pose, die der Marker zum Zeitpunkt des Wechsels vom Tracking- in den Inspektionsmodus hatte, wird zwischengespeichert und als Basis für die Transformationen weiterverwendet.

# 7. Fazit

Dieses Kapitel fasst die Arbeit zusammen und erklärt die geschaffene festgelegten Marker richten, und mit dem dann auf dem Bildschirm sichtbaren 3D-Modell interagieren. Die grundlegenden Transformationen (Rotieren, Verschieben, Skalieren) werden über Bewegungen auf dem Touchscreen auf das Modell angewendet. Außerdem kann der Benutzer das Modell optional auch unabhängig vom Marker betrachten.

Um dem Anwender darüber hinaus Informationen zur Verfügung zu stellen, kann er ein einzelne Teile des Modells berühren. Das ausgewählte Teil wird über eine geringe Translation vom Modell weg hervorgehoben, die passenden Informationen werden in einer Textbox angezeigt.

Daneben wurde mit dem Enumerations-Typ die Grundlage für ein modulares System zum schnellen Austausch von Modelldaten erstellt.

## 7.1. Limitationen

Das Nutzungserlebnis der Anwendung hängt stark vom zugeführten Inhalt ab. Ein schlecht umgesetztes 3D-Modell, unzureichende Informationen oder fehlende Interaktionsflächen des Modells kann die Anwendung alleine nicht aufwiegen.

Der Einfluss auf die AR-Komponente ist begrenzt. Da mit dem QCAR SDK eine proprietäre Lösung eingesetzt wird, hat der Entwickler keinen Zugriff auf Interna des SDK, um gegebenenfalls Änderungen z. B. am zugrunde liegenden

Tracking-Algorithmus durchzuführen. Bezüglich der Unterstützung von unterschiedlichen Smartphones und Tablet-PCs ist man ebenfalls auf den Hersteller angewiesen.

Die Anwendung ist natürlich keine mobile Alternative zu professionellen CAD-Anwendungen. Im Vergleich zu aktuellen stationären Rechnern bieten mobile Geräte weniger Leistung. Dazu sind über den Touchscreen keine hochpräzisen Eingaben möglich, und der Bedienkomfort eines stationären Rechners mit großem Bildschirm, Tastatur und Maus kann nicht erreicht werden.

## 7.2. Erweiterungsmöglichkeiten

Die Informationen zu dem 3D-Modell sind in einer Enumeration zusammengefasst, und daher prinzipiell austauschbar. Aktuell werden die Informationen zur Kompilierungszeit fest in die Anwendung integriert. Um die Modifizierbarkeit der Applikation zu verbessern, könnten die Informationen in eine separate Datei auf dem Gerät ausgelagert werden, die bei Bedarf und ohne Neukompilierung ausgetauscht werden können.

Denkbar wäre auch ein Ansatz, der es ermöglicht, das Modell und die Informationen von einem externen Gerät nachzuladen, z. B. über Near Field Communication (NFC, deutsch: „*Nahfeld-Kommunikation*")[47], WLAN oder Bluetooth. Neben einer zusätzlichen Kommunikationskomponente, wäre für die beiden genannten Möglichkeiten auch eine Standardisierung der Informationsstruktur notwendig, etwa über eine XML-Definition.

Das Accelero- und Gyrometer bleiben in der vorliegenden Implementierung unbenutzt. Alternativ zum Touchscreen könnten alle drei Transformationen auch über über diese Sensoren gesteuert werden. Wird das Gerät über die Querachse zur Seite gekippt, würde das Modell je nach Richtung und Modus nach

---

[47]Eine Technologie zur Datenübertragung über sehr kurze Distanzen von bis zu 4 cm [NFCF]

links oder rechts verschoben bzw. rotiert. Bei einer Neigung über die Längsach-se nach vorne oder hinten würde das Modell verkleinert oder vergrößert. Der Nachteil dabei ist allerdings, dass die Sensoren auch bei normalen Bewegun-gen des Gerätes das 3D-Objekt (ungewollt) transformieren würden. Vor jeder gewollten Transformationen müsste also die Nutzung der Sensoren aktiviert, und danach wieder deaktiviert werden.

Alternativ zu den Lagesensoren kann auch der Multitouch Bildschirm stär-ker genutzt werden. Anstatt aus dem Menü eine Achse auszuwählen, wird die Transformationsachse in Abhängigkeit der Anzahl der Finger auf dem Touch-screen bestimmt. Bei einem Finger könnte das Modell z. B. auf der X-Achse verschoben/rotiert werden, bei zwei Fingern auf der Y-Achse und bei Drei auf der Z-Achse.

Um einzelne Teile am Modell auszutauschen (wie in der Anwendung von Mer-cedes in Abschnitt 3.3.2), müssten die entsprechenden Ersatzteile ebenfalls als 3D-Modell vorliegen. Da mit jPCT der Zugriff auf die einzelnen Teile möglich ist, könnte das auszutauschende Teil während der Laufzeit unsichtbar bzw. das neue Teil sichtbar gemacht werden.

Das Hinzufügen von Notizen für einzelne Bauteile ist ebenfalls denkbar. Über die virtuelle Tastatur, die Android zur Eingabe einblenden kann, könnten in ein Eingabefeld kurze Texte geschrieben und gespeichert werden. Dies wäre etwa bei den in Abschnitt 4.2.3 erwähnten Besprechnungen nützlich. Da mit Android die persistente Speicherung von Daten möglich ist, würden diese In-formationen auch beim Beenden der Anwendung oder Ausschalten des Gerätes nicht verloren gehen.

# A. Anhang

# Verwendete Ressourcen

### LaTeX Vorlage

Diese Arbeit basiert auf der LaTeX Vorlage für Masterarbeiten von Stefan Macke und wurde vom Autor modifiziert.

Lizenz: CC-BY-SA

URL: `http://blog.stefan-macke.com/2009/04/24/latex-vorlage-fuer-meine-masterarbeit-an-der-ohm-hochschule-nuernberg/`

### Marker Textur

Die Textur, die für den Marker verwendet wurde, stammt vom Autor *jaqx-textures.*

Lizenz: CC-BY-NC

URL: `http://jaqx-textures.deviantart.com/art/Tarmac-painted-2-51647406`

### Microsoft Visio Shapes

Für einige Grafiken wurden die 3D-Shapes für Microsoft Visio von Junichi Yoda verwendet.

Lizenz: Nicht angegeben, „You are free to use these files"

URL: `http://www.geocities.jp/newpipingisome/index.html`

Auf der nächsten Seite (Anhang III) befindet sich der oben genannte Marker im DIN A4 Format.

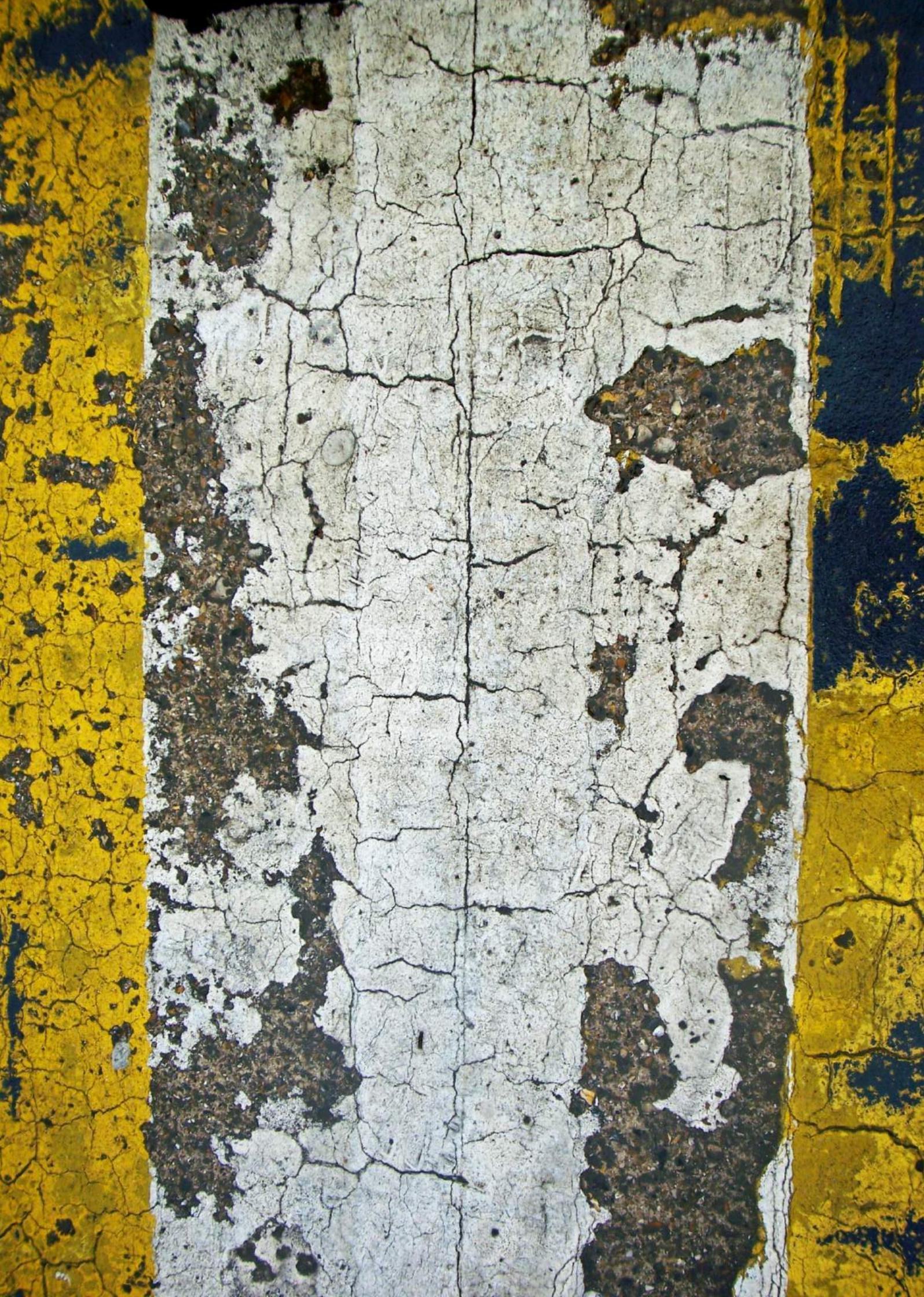

# Quellcode-Struktur der Implementierung

## Paket ARi3D

**ARinteract3D.java** Haupt-Activity der Anwendung, enthält QCAR-Steuer-code und die Callback-Methoden für Touchscreen-Eingaben und Menü.

**DebugLog.java** Teil des QCAR-SDKs. Enthält keine Eigenleistungen.

**QCARSampleGLView.java** Teil des QCAR-SDKs. Enthält keine Eigenleistungen.

**Renderer.java** Implementierung des OpenGL-Renderers. Steuert die Darstellung und Kommunikation mit dem AR-Tracker.

## Paket ARi3D.Helper

**CarParts.java** Enumerations-Typ mit Informationen über das 3D-Modell.

**Helper.java** Beinhaltet verschiedene Hilfs-Methoden, u. a. das Laden von serialisierten 3D-Modellen.

**MessageHandler.java** Vom Renderer zur Kommunikation mit der Haupt-Activity benötigt.

# JNI Klassen

**Ordner QCAR** Header-Dateien des QCAR-SDKs. Enthält keine Eigenleistungen.

**ARi3D.cpp** Hauptklasse des AR-Trackers. Berechnung der Matrizen, wird vom Renderer aufgerufen.

**CubeShaders.h** Teil des QCAR-SDKs. Enthält keine Eigenleistungen.

**SampleMath.cpp** Benötigt für Matrix-Operationen in ARi3D.cpp. Teil des QCAR-SDKs. Enthält keine Eigenleistungen.

**SampleMath.h** Teil des QCAR-SDKs. Enthält keine Eigenleistungen.

**SampleUtils.cpp** Benötigt für Matrix-Rotation, -Translation, -Skalierung. Teil des QCAR-SDKs. Enthält keine Eigenleistungen.

**SampleUtils.h** Teil des QCAR-SDKs. Enthält keine Eigenleistungen.

# Glossar

**C**

**Canny Detektor** Ein Algorithmus aus der digitalen Bildverarbeitung zur Erkennung von Kanten in Bildern. Er wurde 1986 von John Canny entwickelt [Canny86] und besteht aus mehreren Einzelschritten: Glättung des Bildes, Berechnung der Steigung der Kanten, Non-maximum Unterdrückung und Schwellenwert-Hysterese. Das Ergebnis ist ein Binärbild, in dem die Kanten weiß eingezeichnet sind.

**E**

**Enumeration** Kurz Enum, deutsch: „*Aufzählung*". Bezeichnet in Java einen Datentyp, der aus typensicheren Konstanten besteht [Ora].

**O**

**Open Handset Alliance** Eine Gruppe aus 84 Unternehmen aus dem Mobilfunk- und Technologiebereich, die Innovationen im mobilen Sektor fördern wollen, und mit Android eine kostenlose und offene Software-Plattform entwickelt haben [OHA].

**S**

**Serialisierung** Speichern von Daten oder einer Datenstruktur in einer externen Datei. Hier wurde die Modelldatei und die dazugehörige Material/Oberflächendatei vorab mit der Desktop-Version von jPCT in eine einzige Datei serialisiert. jPCT-AE kann die serialisierte Datei schneller einlesen als die beiden separaten Dateien, was zu einer schnelleren Ladezeit beim Programmstart führt.

# Abkürzungsverzeichnis

| | |
|---|---|
| AR | Augmented Reality |
| CAVE | Cave Automatic Virtual Environment |
| CCD | Charged-Coupled Device |
| CV | Computer Vision |
| DMU | Digital Mock-Up |
| FAST | Features from Accelerated Segment Test |
| FFP | Fixed Function Pipeline |
| GLSL | GL Shading Language |
| GPU | Graphics Processing Unit |
| HMD | Head Mounted Display |
| JNI | Java Native Interface |
| NDK | Native Development Kit |
| NFC | Near Field Communication |
| OpenGL-ES | OpenGL for Embedded Systems |
| PTAM | Parallel Tracking and Mapping |
| QCAR | Qualcomm Augmented Reality SDK |
| SAR | Spatial AR |
| SIFT | Scale Invariant Feature Transform |
| SLAM | Simultaneous Localization and Mapping |
| VR | Virtuelle Realität |

# Abbildungsverzeichnis

# Tabellenverzeichnis

# Verzeichnis der Listings

# Literatur

## Literaturverzeichnis

## Online-Quellen

# Bildquellen

# Eigenständigkeitserklärung

Ich, Lennart Brüggemann, versichere hiermit, dass die vorliegende Bachelorarbeit mit dem Thema

*Interaktion mit 3D-Objekten in Augmented Reality Anwendungen*
*auf mobilen Android Geräten*

selbstständig und nur unter Verwendung der angegebenen Quellen und Hilfsmittel verfasst habe. Die Arbeit wurde bisher in gleicher oder ähnlicher Form keiner anderen Prüfungsbehörde vorgelegt.

Berlin, den 14. Februar 2012

—————————————————

Lennart Brüggemann